\documentclass[aps,prd,longbibliography,nofootinbib]{revtex4-2}

\usepackage{graphicx}
\usepackage{dcolumn}
\usepackage{amsmath,amssymb,epsfig,latexsym}
\usepackage{paralist}
\usepackage{comment}
\usepackage{graphicx}
\usepackage{multirow}
\usepackage{color,soul}
\usepackage{suffix}
\usepackage{mathtools}

\usepackage{siunitx}
\usepackage{longtable}
\usepackage{array}
\usepackage{natbib}
\usepackage{hyperref}
\hypersetup{colorlinks=true,allcolors=blue}

\newcommand{\Cov}{\mathbf{C}}
\newcommand{\Cshot}{\mathbf{C}_{\rm shot}}
\newcommand{\Catm}{\mathbf{C}_{\rm atm}}
\newcommand{\Cinst}{\mathbf{C}_{\rm inst}}
\newcommand{\Cosc}{\mathbf{C}_{\rm osc}}
\newcommand{\Cnl}{\mathbf{C}_{\rm nl}}

\usepackage{tikz}
\usetikzlibrary{arrows,positioning,calc}

\begin{document}

\title{High-Precision Amplitude-Modulated Continuous-Wave Lunar Laser Ranging}

\author{Slava G. Turyshev}   

\affiliation{ 
Jet Propulsion Laboratory, California Institute of Technology,\\
4800 Oak Grove Drive, Pasadena, CA 91109-0899, USA
}%

\date{\today}

\begin{abstract}
Lunar laser ranging (LLR) currently delivers mm-class tests of relativistic gravity and the lunar interior, but further gains are limited by photon-starved pulsed systems, array-induced pulse broadening, and atmospheric variability. This paper develops the metrology and covariance layer for high-power amplitude-modulated continuous-wave (AM--CW) LLR. The optical link budget and kW-class CW architecture are taken from the companion high-power CW LLR analysis; here the focus is on RF-envelope phase observables, multi-tone ambiguity removal, range and range-rate estimators, detector requirements, Doppler derotation, and observation-level covariances. For a GHz-class precision tone, \(c/(4\pi f_m)=2.38567\times10^{-2}~\mathrm{m\,rad^{-1}}\), so \(0.10~\mathrm{mm}\) photon-limited range precision requires \({\rm SNR}_{\tt AM}\simeq240\). With detected photon rates appropriate to a 1~kW, 1064~nm transmitter on a 1--2~m class telescope ranging to 10~cm corner-cube retroreflectors, the \(T\simeq100~\mathrm{s}\) photon-statistical range floor is \(0.08\)--\(0.14~\mathrm{mm}\) in a generic high-power case, \((3\)--\(6)\times10^{-5}~\mathrm{m}\) in a dedicated AM--CW case, and \(\lesssim3\times10^{-5}~\mathrm{m}\) in a photon-rich case. With representative residual atmosphere and instrument allocations, a dedicated station can plausibly reach \(\sim0.08~\mathrm{mm}\) absolute range precision under favorable conditions. Range-rate precision below \(1~\mu\mathrm{m\,s^{-1}}\) requires several-hundred-second windows, or shorter windows only in photon-rich operation. Differential LLR between nearby lunar reflectors suppresses common-mode station and atmospheric terms, but it cannot suppress independent photon noise. For equal links,
\(\sigma_{\Delta R,\rm shot}=\sqrt{2}\,\sigma_{R,\rm shot}\). Thus the \(20~\mu\mathrm{m}\) differential level is a stretch goal requiring higher detected flux, longer integration, or both. Robust design bands are \(\sim45\)--\(90~\mu\mathrm{m}\) for the dedicated AM--CW case and \(\sim35\)--\(60~\mu\mathrm{m}\) in photon-rich excellent-seeing operation. The resulting requirements on link SNR, Doppler derotation, detector mode,
instrument PSD/Allan stability, oscillator slew, multi-tone nonlinearity, and differential CONOPS are presented.

\end{abstract}

\maketitle


\section{Introduction}
\label{sec:scope}

Since its inception in 1969, lunar laser ranging (LLR) has achieved millimeter-level precision in measuring the distance between Earth-based observatories and the corner-cube retroreflectors (CCRs) on the Moon~\cite{Dickey1994Science,Murphy2008PASP,Murphy2013RoPP,Williams2004PRL,Williams2012CQG}. This progress has enabled a suite of high-precision tests of relativistic gravity, including tests of the equivalence principle, constraints on a possible variation of $G$, geodetic precession, and detailed studies of the deep lunar interior~\cite{Williams2004PRL,Williams2012CQG}. In this classical pulsed architecture, the ranging observable is the distribution of photon arrival times referenced to an internal clock, and the range estimate is obtained by fitting this distribution after applying corrections for relativistic light time, station motion, and atmospheric delay. However, further improvements are increasingly limited by photon-starved statistics, array-induced pulse broadening, and two-way atmospheric variability, specifically:
\begin{itemize}
\item \emph{photon statistics:} the number of detected photons per normal point is modest, and further gains by increasing
peak power are constrained by eye safety, nonlinearity, and cost;
\item \emph{array--induced pulse spreading:} large corner--cube arrays produce return pulses broadened by array geometry
and librations, limiting the utility of higher single--photon timing precision;
\item \emph{atmosphere and station:} two--way atmospheric delay variations and station metrology become non--negligible
once intrinsic photon noise is driven below the millimeter level.
\end{itemize}

Recently, we analyzed a high-power CW LLR link for next-generation 10~cm lunar CCRs, including photon-return rates, optical throughput, background, and first-order station error budgets~\cite{Turyshev:CW-LLR:2025,Turyshev-CCR:2025}.
The present paper should be read as the metrology and covariance continuation of that work, not as a second link-budget paper. We adopt the same 1~kW, 1064~nm, 10~cm-CCR architecture as a baseline and focus on the AM--CW measurement layer: RF-envelope phase observables, ambiguity removal, range and range-rate estimators, detector requirements, Doppler derotation, and the covariance matrix used by global LLR analyses.

AM--CW phase ranging is a mature metrological technique. Optical intensity-modulation distance meters have demonstrated high-resolution microwave phase ranging, including modulation near 28~GHz~\cite{Fujima1998MST}; modern free-space electro-optical distance meters have demonstrated
sub-mm performance over multi-km terrestrial paths with atmospheric compensation~\cite{Guillory2024MST}. The novelty here is thus not the basic AM--CW principle, but its application to the lunar link: a \(\simeq 2.56~\mathrm{s}\) round trip, fW-level received optical powers, GHz-class RF-envelope phase recovery, lunar Doppler derotation, multi-tone integer ambiguity resolution, and a covariance model suitable for
sub-mm LLR normal points.

In the AM--CW architecture an RF tone at frequency \(f_m\) modulates the optical power envelope, the lunar CCR returns a delayed and attenuated copy of that envelope, and the station estimates the returned RF phase relative to a maser-referenced local oscillator. The mean unwrapped phase gives the two-way range observable \(R_{2\mathrm{w}}\), while the phase slope gives the one-way line-of-sight range-rate \(v_r\), equivalently the Doppler shift of the modulation envelope.

Differential LLR is obtained by interleaving measurements between two lunar
CCRs that are separated on the lunar surface but close on the sky. The angular
separation corresponding to a surface baseline \(B_{\rm AB}\) is
\begin{equation}
  \theta_{\rm AB}\simeq \frac{B_{\rm AB}}{L_{\rm EM}}
  =
  0.149^\circ
  \left(\frac{B_{\rm AB}}{10^3~\mathrm{km}}\right)
  \left(\frac{3.84\times 10^8~\mathrm{m}}{L_{\rm EM}}\right).
\end{equation}
Thus \(\theta_{\rm AB}\lesssim 0.1^\circ\) corresponds to
\(B_{\rm AB}\lesssim 670~\mathrm{km}\) at mean lunar distance, whereas a
\(10^3~\mathrm{km}\) baseline corresponds to
\(\theta_{\rm AB}\simeq 0.15^\circ\). The differential observables
\((\Delta R_{2\mathrm{w}},\Delta v_r)\) suppress common-mode station terms, but their precision remains bounded by independent photon noise from the two returns and by residual differential atmosphere.

To enable AM--CW LLR, the key hardware element is an RF phase--metrology chain, phase--locked to an ultra-stable frequency reference and used to modulate a kW-class, 1064\,nm CW laser, with the lunar CCRs supplying a weak, delayed replica of the imposed RF modulation. Compared to pulsed LLR, this architecture supports sustained coherent integration of the RF tone(s), greatly reduced effective
detection bandwidth via narrowband digital lock-in detection, and routine use of GHz-class modulation frequencies. Taken together, these features open a realistic path to sub--0.1\,mm two--way range precision and robust, high-sensitivity range--rate measurements on $T\sim 10^2$--$10^3$\,s~windows.

From a science perspective, sub--0.1\,mm absolute LLR and tens--of--$\mu$m differential LLR open several new
regimes:
\begin{itemize}
\item measurement of ultra--slow changes in the lunar tidal dissipation and Love numbers on decadal time scales;
\item detection of small shifts in libration amplitudes and precession that may indicate core--mantle coupling or exotic
interior structure; 
\item sensitivity to very low--frequency ($\mu$Hz) gravitational waves, which imprint characteristic, slowly varying
signatures on the Earth--Moon separation;
\item improved constraints on higher--order relativistic effects in the Earth--Moon system, including post--Newtonian
parameters and possible deviations from general relativity at long range.
\end{itemize}

As a result, a highly precise AM--CW implementation of LLR, operated in both absolute and differential modes, would open a qualitatively new regime for tests of relativistic gravity, lunar interior structure, and long-term orbital and rotational dynamics \cite{dLLR-BPS-decadal,Turyshev:CW-LLR:2025}. In particular, dLLR provides enhanced sensitivity to spatially correlated signatures—such as deep-mantle tidal response, core--mantle coupling, and possible violations of the equivalence principle—by exploiting simultaneous measurements to multiple reflectors with strongly suppressed common-mode noise \cite{Zhang2022DLLR,Zhang2024LLR_DLLR,Zhang2023DLLRThesis}. The recent detection of a dense solid inner core and associated mantle overturn in the Moon further sharpens the need for such high-precision constraints on the deep interior \cite{Briaud2023NatureCore}. At the same time, this level of ranging accuracy enables searches for stochastic and deterministic gravitational-wave signals in the $\mu$Hz band \cite{BlasJenkins2022PRL,BlasJenkins2022PRD,BlasJenkinsTuryshev2024FunPAG}.

LLR already sits near the top of the gravity--test hierarchy; the question this paper addresses is what hardware,
measurement models, and operating procedures are needed to push the photon--rich AM--CW approach into a regime
where $\sim 0.1$\,mm normal points and high--precision range--rate are technically realistic.

Our goal here is to determine what station hardware, measurement models, and observing strategy are required to achieve sub-0.1\,mm two-way range precision and $(0.1$--$1)\,\mu\mathrm{m\,s^{-1}}$ range-rate sensitivity over $\sim10^2$--$10^3$\,s, and to translate these requirements into a quantitative error budget and concept of operations (CONOPS).

A detailed link budget and hardware feasibility study for high-power CW LLR, including realistic photon return rates to 10 cm CCRs on a 1 m class telescope, is developed in \cite{Turyshev:CW-LLR:2025}. Here we adopt that architecture (1~kW at 1064~nm on a 1~m aperture interrogating 10~cm CCRs) as a baseline and focus on the metrology side: a unified AM--CW phase model, a covariance description suitable for parameter estimation, and system-level error budgets that map directly onto the observables $\big(R, v_{r}, \Delta R,\Delta v_{r}\big)$ and onto hardware and CONOPS requirements.

In particular, the present work introduces: (i) a unified AM--CW phase--measurement model and joint covariance ${\bf C}(T)$ for the observable vector $y(T)=\bigl(R_{2\mathrm{w}},v_r, \Delta R_{2\mathrm{w}},\Delta v_r\bigr)^{\rm T}$; (ii) an explicit multi-tone synthetic-wavelength ambiguity-removal scheme with quantitative nonlinearity constraints on $\Cnl(T)$; and (iii) a system-level error budget and observatory concept of operations that map the photon, atmospheric, instrumental, oscillator, and nonlinearity contributions directly into hardware and CONOPS requirements for next-generation AM--CW LLR stations.

Conceptually, the core object of this paper is the observation covariance matrix ${\bf C}(T)$ of the AM--CW range and range-rate estimators over an integration window $T$. This covariance is decomposed into photon, atmospheric, instrumental, oscillator, and nonlinearity contributions. The station hardware and operational concept can then be designed so that each subsystem keeps its contribution to ${\bf C}(T)$  within its allocated variance.

This paper is organized as follows: Section~\ref{sec:model} develops a unified observation model for the AM envelope phase, defining the two-way range $\widehat{R}_{2{\rm w}}(T)$ and one-way range--rate $\widehat{v}_r(T)$ together with their differential counterparts and the covariance description $\mathbf{C}(T)$. Section~\ref{sec:unwrap} discusses multi-tone ambiguity resolution via synthetic wavelengths. Section~\ref{sec:snr} links photon-counting statistics and the optical link budget to the shot-noise contribution $\mathbf{C}_{\rm shot}(T)$ of the observation covariance. Section~\ref{sec:errors} quantifies the atmospheric and instrumental contributions $\mathbf{C}_{\rm atm}(T)$ and $\mathbf{C}_{\rm inst}(T)$ as functions of integration time $T$ and angular separation~$\theta$. Section~\ref{sec:gates} formulates implementation gates that ensure each hardware and environmental contribution remains within its allocated error budget, and Section~\ref{sec:perf} summarizes the resulting performance bands for absolute and differential range and range--rate. Section~\ref{sec:hw} describes a representative high-power AM--CW LLR station, including the common hardware stack, facility-level infrastructure, and a concept of operations tailored to multi-tone phase metrology. In Section~\ref{sec:concl} we summarize results obtained and conclude.  

\section{AM phase model and observables}
\label{sec:model}

The total one--way optical path length (in meters) between an Earth-based LLR station and a CCR on the moon is
\begin{equation}
  R_{\rm tot}(t) = R_{\rm geom}(t;\boldsymbol{\theta}) + R_{\rm atm}(t) + R_{\rm inst}(t),
  \label{eq:Rtot}
\end{equation}
where $R_{\rm geom}$ includes relativistic light--time, station position and motion, Earth orientation, CCR position and motion with respect to the luni-centric celestial reference system (LCRS) \cite{Turyshev:2025,Turyshev-scales:2025} and lunar ephemeris (see modeling details in \cite{Williams2004PRL,Williams2012CQG}.) The atmospheric term $R_{\rm atm}$ represents the neutral delay mapped to path length via standard refractive index models and mapping functions\footnote{At typical mid--latitude observatories the mean zenith hydrostatic delay at near--IR wavelengths is $\sim2~\mathrm{m}$ in the one--way optical path. Modern surface--pressure--driven mapping functions and refractivity models remove this static component at the $\lesssim 10^{-4}$ level  \cite{Ciddor1996AO,MendesPavlis2004GRL,
Niell1996JGR,Boehm2006GRL}, so that the contribution of $R_{\rm atm}(t)$ to the covariance matrix $C_{ij}(T)$ is set by the residual, time--varying wet delay and turbulent fluctuations rather than by the absolute $2~\mathrm{m}$ column delay, as quantified in Sec.~\ref{sec:atmo} and Appendix~\ref{app:atm_turbulence}.
} \cite{Ciddor1996AO,MendesPavlis2004GRL,Niell1996JGR,Boehm2006GRL}. The instrumental term $R_{\rm inst}$ collects the internal optical and RF path contributions that are not part of the formal geometric model and are monitored by the internal reference.

For a single--tone AM--CW transmitter, the optical power at the telescope aperture may be written as
\begin{equation}
P_{\rm tx}(t) = P_0\Big(1 + a_m \cos(2\pi f_m t)\Big), \qquad 0 < a_m \le 1,
  \label{eq:transm}
\end{equation}
where $P_0$ is the mean optical power and \(a_m\) is the dimensionless amplitude--modulation index. 

After propagation to the Moon, reflection from a CCR, and return to the receiver, the modulation is delayed by the total round--trip light time and attenuated by the link. Neglecting scintillation--induced amplitude fluctuations for the moment, the received power at the detector can be written, to first order in \(a_m\), as
\begin{equation}
P_{\rm rx}(t) = P_1\Big(1 + a_m \cos\bigl(2\pi f_m (t - \tau(t))\bigr)\Big), \qquad
\tau(t) = \frac{2R_{\rm tot}(t)}{c},
\label{eq:Prx}
\end{equation}
where $P_1$ collects the geometric loss, atmospheric transmission, and reflector response, and $\tau(t)$ is the round--trip light time corresponding to the total station--to--reflector path $R_{\rm tot}(t)$ of Eq.~\eqref{eq:Rtot}. (In the photon--counting description developed in Sec.~\ref{sec:snr}, the same mean received power may be viewed as the optical power corresponding to the detected photon rate, so that $P_1 \simeq \dot N_\gamma E_\gamma$ with $\dot N_\gamma$ given by the link model of Eq.~\eqref{eq:link} and $E_\gamma$ defined there, also see \cite{Turyshev:CW-LLR:2025}.)

After direct square--law detection and narrowband filtering around $f_m$, the photocurrent on a given tone can be modeled as
\begin{equation}
  i(t) = I_0 + I_1 \cos \bigl(2\pi f_m t - \phi(t)\bigr) + n(t),
  \label{eq:iRF}
\end{equation}
where \(I_0\) is the mean photocurrent, \(I_1 \propto a_m P_1\) is the amplitude of the modulation component,  \(n(t)\) is dominated by shot noise and residual background, and \(\phi(t)\) is the modulation--envelope phase given as
\begin{equation}
  \phi(t) = \phi_0 +
  \frac{4\pi f_m(t)}{c}\,R_{\rm tot}(t) + \epsilon_\phi(t),
  \label{eq:phiRF}
\end{equation}
with $\epsilon_\phi$ the phase noise from photon statistics, electronics, and fast turbulence. In the high--SNR limit, efficient estimators satisfy $\sigma_\phi^2 \simeq 1/{\rm SNR}_{\tt AM}^2$ [rad$^2$] for the tone phase~\cite{GoodmanStatOptics,KayEstimation,RifeBoorstyn1974}.

The two primary observables are:
\begin{itemize}
\item two--way range $R_{2{\rm w}}$ from the mean phase $\bar\phi$,
\item one--way range--rate $v_r$ from the phase slope $b = d\phi/dt$.
\end{itemize}
Differential observables, formed between two CCRs, further suppress station--level and atmospheric contributions.

\subsection{Science-driven metrology requirements}
\label{sec:reqs}

Building on the high-power CW link and reflector feasibility analysis of \cite{Turyshev:CW-LLR:2025}, the present work focuses on the metrology side of the problem: a unified AM--CW phase model, a covariance framework suitable for global parameter estimation, and quantitative station-level error budgets that map directly onto sub-mm normal-point observables and subsystem requirements. Our objective is to translate the science-driven precision goals into requirements on the station hardware, measurement model, and observing strategy.

For later parameter estimation it is useful to write a linearized observation model for the two--way range,\footnote{We adopt the standard LLR convention that the estimated ``two--way range'' $R_{2\mathrm{w}}(T)$ is defined as the \emph{one--way} geometric Earth--Moon distance inferred from the round--trip light time. Thus the modeled optical path entering the light--time calculation is
$2R_{\rm tot}(t)$, while the estimator $\widehat{R}_{2\mathrm{w}}(T)$ defined in Eq.~\eqref{eq:Robs} returns a quantity numerically equal to the time--averaged $R_{\rm tot}(t)$ over the analysis window~$T$.}
\begin{equation}
  \delta R_{2{\rm w}}(t_k) =
  \sum_j \frac{\partial R_{2{\rm w}}}{\partial \theta_j}(t_k)\,\delta\theta_j
  + \epsilon_R(t_k),
\label{eq:linobs}
\end{equation}
where $\boldsymbol{\theta}$ is the set of dynamical and relativistic parameters (e.g.\ lunar Love numbers, tidal $Q$,
equivalence--principle parameters, and a possible $\dot G/G$), and $\epsilon_R$ denotes the residual measurement noise
with variance $\sigma_R^2(T)$ on a window of duration $T$.  For a single parameter $\theta_j$ that is not strongly
correlated with others, the formal uncertainty from $N$ normal points is approximately
\begin{equation}
  \sigma^2(\theta_j) \simeq
  \left[
    \sum_{k=1}^{N}
      \frac{1}{\sigma_R^2(T_k)}
      \left(
        \frac{\partial R_{2{\rm w}}}{\partial \theta_j}(t_k)
      \right)^2
  \right]^{-1}.
\label{eq:fisher}
\end{equation}

Eqs.~\eqref{eq:linobs}--\eqref{eq:fisher} make explicit the otherwise qualitative statement that reducing the single-station two-way uncertainty from the current millimeter level to the sub--$0.1$~mm regime improves sensitivity to the underlying physics parameters by roughly an order of magnitude, all else equal. In the unified AM--CW observation model developed in Sec.~\ref{sec:model}, the window-dependent variance $\sigma_R^2(T)$ entering these expressions is simply the $(1,1)$ element of the joint observation covariance matrix $\mathbf{C}(T)$ of the range and range-rate estimators [see Eq.~\eqref{eq:covsplit}]. Throughout the rest of the paper we treat $\mathbf{C}(T)$ as the central object that links hardware and environmental noise to the science parameters~$\boldsymbol{\theta}$.

The same observation-level covariance applies to the full AM--CW observable set. In the scalar case of Eq.~\eqref{eq:linobs}, the variance \(\sigma_R^2(T)\) is the \((1,1)\) element \(C_{11}(T)\) of \(\mathbf{C}(T)\). In the four-observable case, the data vector for a window \(T_k\) is \(\mathbf{y}(T_k)\) in Eq.~\eqref{eq:yvec-global}, with covariance \(\mathbf{C}(T_k)\); the Fisher matrix is then built from all four components \((R_{2\mathrm{w}},v_r,\Delta R_{2\mathrm{w}},\Delta v_r)\) and the corresponding rows of the design matrix \(\mathbf{H}(T_k)\) in Eq.~\eqref{eq:global-observation}. Absolute range at the \(\sim0.1~\mathrm{mm}\) level improves the observation-level sensitivity to post-Newtonian parameters, equivalence-principle signals, \(\dot G/G\), tidal parameters, and lunar-interior signatures, but the final science gain is set by the full global covariance, including parameter correlations and systematic model errors. Differential observables between nearby next generation CCRs are best interpreted as high-precision probes of spatial gradients in the lunar response and as controls on station and atmospheric systematics, rather than as standalone guarantees of a particular parameter accuracy.

As a displacement scale, a very low-frequency
\((f\ll 1/\tau_{2{\rm w}})\) gravitational-wave perturbation may be represented as a slowly varying effective displacement of the Earth--Moon separation \cite{Turyshev-SGWB_LLR:2026}. A useful order-of-magnitude conversion is
\begin{equation}
  \delta R_{2{\rm w}}(t)
  \sim {\cal O}(1)\,\frac{hL_{\rm EM}}{2},
  \qquad
  h_{\rm disp}\sim \frac{2\sigma_R}{L_{\rm EM}} .
\end{equation}
With \(L_{\rm EM}\simeq 3.84\times 10^8~\mathrm{m}\),
\[
  h_{\rm disp}\simeq
  5.2\times 10^{-13}
  \left(\frac{\sigma_R}{0.1~\mathrm{mm}}\right).
\]
This is a displacement-equivalent scale, not a detection threshold. A detection calculation must include the binary response function, antenna pattern, temporal filtering, colored multi-epoch covariance, and degeneracies with ephemeris, station, geophysical, and lunar-interior parameters. Similar
caution applies to tidal and libration signatures: the final science gain is set by the full global covariance, not by the single-window range variance alone.\footnote{Baseline constants: speed of light $c=299\,792\,458$\,m\,s$^{-1}$, Earth-Moon distance $L_{\rm EM}\simeq 3.84\times 10^8$\,m. Two--way light time near opposition is $\langle \tau\rangle\approx 2.56$\,s. We assume standard relativistic light--time modeling for $R_{\rm geom}(t;\boldsymbol{\theta})$ \cite{Williams2004PRL,Williams2012CQG} and standard models for atmospheric refractivity and mapping \cite{Ciddor1996AO,MendesPavlis2004GRL,Niell1996JGR,Boehm2006GRL}.}

\subsection{Unified phase--measurement model}

After subtraction of the internal reference phase $\phi_{\rm inst}(t)$, the phase samples entering the estimators may be
written as
\begin{equation}
  \tilde\phi(t_k) =
  \frac{4\pi f_m^\ast}{c}\,R_{\rm tot}(t_k)
  + \delta\phi_{\rm slew}(t_k)
  + \delta\phi_{\rm osc}(t_k)
  + \delta\phi_{\rm nl}(t_k)
  + n_\phi(t_k),
\label{eq:meas}
\end{equation}
where $f_m^\ast$ is the window--averaged modulation frequency, $\delta\phi_{\rm slew}(t_k)$ is the deterministic contribution from modulation-frequency drift across the round trip, $\delta\phi_{\rm osc}(t_k)$ collects residual stochastic oscillator phase noise after slew removal, $\delta\phi_{\rm nl}(t_k)$ collects residual hardware nonlinearities (e.g.\ AM--to--PM conversion, frequency-dependent RF delays, small differences between the internal and lunar paths), and $n_\phi(t_k)$ is a zero-mean stochastic term dominated
by photon statistics and fast turbulence. In the high--SNR regime of interest here, $n_\phi$ is well approximated as white with variance $\sigma_\phi^2 \simeq 1/{\rm SNR}_{\tt AM}^2$ (rad$^2$) over the analysis bandwidth, where ${\rm SNR}_{\tt AM}(T)$ denotes the lock--in SNR accumulated over the window~$T$ as introduced in Sec.~\ref{sec:snr}  by Eq.~(\ref{eq:lockin_snr}).

Decomposition of the phase (\ref{eq:meas})  into the deterministic path term, the known modulation-frequency
slew, and the residual stochastic processes makes explicit that all hardware and environmental effects enter the AM--CW LLR problem only through $R_{\rm tot}(t)$, the deterministic slew, and the residual phase fluctuations $\delta\phi_{\rm osc}(t)$, $\delta\phi_{\rm nl}(t)$, and $n_\phi(t)$. Because the range and range-rate estimators are linear functionals of the
sequence $\{\tilde\phi(t_k)\}$, on a single reflector and tone, the jointly estimated observables over a window $T$ are collected in\footnote{In what follows, we reserve the ``hat'' notation (e.g.\ $\widehat{R}_{2{\rm w}}$) for explicit estimator definitions such as Eqs.~\eqref{eq:Robs}--\eqref{eq:vrob}. When discussing the observables as inputs to the global parameter--estimation problem, we drop the hats and use $R_{2\mathrm{w}}(T)$, $v_r(T)$, $\Delta R_{2\mathrm{w}}(T)$, and $\Delta v_r(T)$ to denote the corresponding estimated quantities.}
\begin{equation}
  \mathbf{y}(T) \equiv
  \begin{bmatrix}
    \widehat{R}_{2{\rm w}}(T) \\
    \widehat{v}_r(T)
  \end{bmatrix},
\end{equation}
so that
\begin{equation}
  \mathbf{y}(T) =
  \mathbf{h}\big(R_{\rm tot};T\big)
  + \boldsymbol{\eta}(T),
  \label{eq:measvec}
\end{equation}
where  $\mathbf{h}$ is the deterministic mapping implied by Eqs.~(\ref{eq:Rtot}), (\ref{eq:Robs}), and (\ref{eq:vrob}),
and $\boldsymbol{\eta}(T)$ are fully characterized, at the Gaussian level, by their covariance matrix ${\bf C}(T)$ via a zero--mean random vector with covariance
\begin{eqnarray}
  \Cov(T) &\equiv&
  {\rm Cov}\big[\boldsymbol{\eta}(T)\big] \equiv
  \left\langle
  \bigl(\boldsymbol{\eta}(T) - \langle \boldsymbol{\eta}(T)\rangle\bigr)
  \bigl(\boldsymbol{\eta}(T) - \langle \boldsymbol{\eta}(T)\rangle\bigr)^{\!\mathsf{T}}
  \right\rangle =\nonumber\\
  &=& \Cshot(T)
   + \Catm(T)
   + \Cinst(T)
   + \Cosc(T)
   + \Cnl(T),
\label{eq:covsplit}
\end{eqnarray}
where ${\bf C}_{\rm shot}(T)$ is set by photon statistics (\ref{eq:shot}) and the sampling model (\ref{eq:lsq}), ${\bf C}_{\rm atm}(T)$ captures atmospheric turbulence and mapping-function residuals, ${\bf C}_{\rm inst}(T)$ collects bench, optical, RF, and internal-reference path-length residuals, ${\bf C}_{\rm osc}(T)$ is determined by the Allan deviation of the frequency reference, and ${\bf C}_{\rm nl}(T)$ represents coherent hardware nonlinearities, including residual conversion of amplitude modulation into phase errors (AM--to--PM conversion) and frequency-dependent path delays that differ between tones.

Subsequent sections assign quantitative allocations to each of these terms: ${\bf C}_{\rm shot}(T)$ via the lock-in SNR ${\rm SNR}_{\tt AM}(T)$ and the Cramér--Rao bounds for phase and slope estimators, ${\bf C}_{\rm atm}(T)$ via Kolmogorov turbulence scalings and mapping functions, ${\bf C}_{\rm inst}(T)$ via internal metrology and bench stability models, ${\bf C}_{\rm osc}(T)$ via the measured Allan deviation of the maser, and ${\bf C}_{\rm nl}(T)$ via multi-tone synthetic-wavelength and AM--to--PM constraints. The absolute range precision $\sigma_R^2(T)$ entering the science model is simply the $(1,1)$ element of ${\bf C}(T)$, and the quantitative error budget is expressed as allocations to the individual covariance terms.\footnote{For clarity, we will often refer to the covariance elements as
\begin{equation}
  \Cov(T) \equiv
  \begin{pmatrix}
    C_{11}(T) & C_{12}(T) \\
    C_{21}(T) & C_{22}(T)
  \end{pmatrix}
  =
  \begin{pmatrix}
    \sigma_R^2(T) & \mathrm{cov}\bigl(R,v_r;T\bigr) \\
    \mathrm{cov}\bigl(R,v_r;T\bigr) & \sigma_{v_r}^2(T)
  \end{pmatrix},
\end{equation}
with the understanding that, when differential observables are included, the same notation is promoted to the full four--observable vector
\((R_{2\mathrm{w}}, v_r, \Delta R_{2\mathrm{w}}, \Delta v_r)\), so that the \(2\times 2\) form written above is just the upper--left block of the \(4\times 4\) covariance matrix used later in the error--budget and parameter--estimation discussions.} 
Eq.~(\ref{eq:covsplit}) provides a compact observation model that can be used directly in Kalman filtering or global parameter estimation. Atmospheric and turbulence models follow standard treatments in
Refs.~\cite{Ciddor1996AO,MendesPavlis2004GRL,Niell1996JGR,Boehm2006GRL,
Tatarskii1961,AndrewsPhillips2005,Roddier1981}, while the photon and estimator statistics are based on Refs.~\cite{GoodmanStatOptics,KayEstimation,RifeBoorstyn1974}.

More generally, when differential observables between two CCRs are formed, each analysis window $T_k$ yields a 4-component vector of observables
\begin{equation}
  \mathbf{y}(T_k) \equiv
  \Big(R_{2\mathrm{w}}(T_k),\,v_r(T_k),\,
        \Delta R_{2\mathrm{w}}(T_k),\,\Delta v_r(T_k)\Big)^{\mathsf{T}}.
  \label{eq:yvec-global}
\end{equation}
Operationally, $R_{2\mathrm{w}}(T)$ and $v_r(T)$ are the primary carriers of global signatures tied to the overall Earth--Moon orbit (EP tests, $\dot G/G$, PPN parameters, and very low-frequency gravitational waves), whereas $\Delta R_{2\mathrm{w}}(T)$ and $\Delta v_r(T)$ are optimally matched to spatially structured and dissipative interior signals (local tidal loading, regional Love-number variations, and time-varying core--mantle coupling). In practice, global analyses will use the full four-component vector $\mathbf{y}(T_k)$ with observable weights set by the corresponding diagonal entries of $\mathbf{C}(T_k)$ in Table~\ref{tab:summary-2w}.

Linearizing about a reference solution gives
\begin{equation}
  \delta\mathbf{y}(T_k)
  = \mathbf{H}(T_k)\,\delta\boldsymbol{\theta}
    + \boldsymbol{\eta}(T_k),
  \label{eq:global-observation}
\end{equation}
where the rows of $\mathbf{H}(T_k)$ are the partial derivatives of the modeled observables with respect to
$\boldsymbol{\theta}$, evaluated from the same dynamical model used in current LLR analyses, and $\boldsymbol{\eta}(T_k)$ is a zero-mean noise vector with covariance matrix
\begin{equation}
  \mathbf{C}(T_k) \equiv
  \operatorname{Cov}\bigl[\boldsymbol{\eta}(T_k)\bigr],
  \label{eq:C-decomp-global}
\end{equation}
whose decomposition into photon, atmospheric, instrumental, oscillator, and nonlinearity contributions is given by Eq.~\eqref{eq:covsplit}.  The scalar range variance $\sigma_R^2(T_k)$ that appears in Eqs.~\eqref{eq:linobs}--\eqref{eq:fisher} is simply the $(1,1)$ element $C_{11}(T_k)$ of this matrix.\footnote{In what follows we denote the elements of this matrix by $C_{ij}(T)$, so that $C_{11}(T)$ and $C_{22}(T)$ correspond to the variances of $R_{2w}(T)$ and $v_r(T)$, respectively, and the off-diagonal terms capture their covariance.} Eq.~\eqref{eq:global-observation} can then be used directly in least-squares or Kalman-filter estimators
in place of the traditional single-observable LLR model.

Below, each contribution in (\ref{eq:covsplit}) is given a quantitative allocation and linked to specific hardware or environmental controls. The photon term ${\bf C}_{\mathrm{shot}}(T)$ is set by the lock--in signal--to--noise ratio ${\rm SNR}_{\tt AM}(T)$ and sampling model through (\ref{eq:shot}) and (\ref{eq:lsq}). Atmospheric fluctuations enter through ${\bf C}_{\mathrm{atm}}(T)$, whose $(1,1)$ element $\sigma_{R,\mathrm{atm}}^2(T)$ is modeled using Kolmogorov turbulence scalings in Sec.~\ref{sec:atmo}. Residual instrumental effects---including internal metrology, bench stability, and RF/optical delays---are captured in ${\bf C}_{\mathrm{inst}}(T)$ and budgeted in Sec.~\ref{sec:instrum-er}. The oscillator term ${\bf C}_{\mathrm{osc}}(T)$ is bounded using the Allan deviation of the frequency reference and the slew correction model of Sec.~\ref{sec:slew}, while ${\bf C}_{\mathrm{nl}}(T)$ represents coherent phase errors from AM--to--PM conversion and multi-tone nonlinearity, constrained by the synthetic wavelength requirements in Sec.~\ref{sec:unwrap}.  The absolute range budget in Eq.~(\ref{eq:budget}) is simply the $(1,1)$ element of Eq.~(\ref{eq:covsplit}) with these allocations.

In parallel with these stochastic, station--level contributions, the deterministic geometric term $R_{\mathrm{geom}}(t;\boldsymbol{\theta})$ collects global modeling ingredients such as station coordinates and velocities, Earth orientation and nutation, lunar librations and ephemerides, reflector locations and orientations, and relativistic light--time corrections \cite{Williams2004IJMPD,Williams2009IJMPD}. Uncertainties in these quantities are represented in the parameter vector $\boldsymbol{\theta}$ and enter the global least--squares or Kalman--filter solution through the sensitivity matrix $H(T_k)$ and the associated parameter covariance, rather than through the per--window noise covariance $C_{ij}(T_k)$. The precision budgets and implementation gates developed below therefore focus on the incremental photon, atmospheric, and instrumental terms introduced by the AM--CW architecture, while assuming corner--cube designs and deployment strategies for which array--induced pulse spreading, thermoelastic deformations, and related reflector--level systematics remain below the atmospheric and bench allocations on the $T \sim 10$--$100\,$s windows considered here.

\subsection{Range and range--rate} 
\label{sec:obs}

Over an integration window of duration $T$, the estimate for the range is given as 
\begin{equation}
  \widehat{R}_{2{\rm w}}(T) =
  \frac{c}{4\pi f_m^\ast}\,
  \Big(\bar\phi(T)-\phi_0+2\pi N\Big), \qquad N \in \mathbb{Z},
  \label{eq:Robs}
\end{equation}
with $f_m^\ast$ the recorded average modulation frequency on that window and $\bar\phi$ the internal--reference--corrected mean phase. Here and throughout we define $R_{2\mathrm{w}}$ as the one--way geometric Earth--Moon distance inferred from the round--trip light time, so that $\tau(t)=2R_{\rm tot}(t)/c$ (as in Eq.~(\ref{eq:Prx})) and $R_{2\mathrm{w}} = R_{\rm tot}$. This convention matches the usual LLR range observable, in which $c\tau/2$ is taken as the range.

Unwrapping $\phi(t)$ over the window and fitting a slope gives
\begin{equation}
  \widehat{v}_r(T) =
  \frac{c}{4\pi f_m^\ast}\,\widehat{b}(T),
  \qquad
  \widehat{b} = \frac{d\phi}{dt}\ \text{(slope fit)},
  \label{eq:vrob}
\end{equation}
for the one--way line--of--sight range--rate. 

The range-rate observable may also be viewed as the Doppler shift of the modulation envelope. For constant transmitted modulation frequency and \(\tau(t)=2R_{\rm tot}(t)/c\), the received envelope contains \(\cos[2\pi f_m(t-\tau(t))]\), whose instantaneous frequency is
\begin{equation}
  f_{\rm rx}(t)
  =
  f_m\left[1-\dot{\tau}(t)\right]
  \simeq
  f_m-\frac{2f_m}{c}\,v_r(t).
\label{eq:am-doppler}
\end{equation}
Thus
\begin{equation}
  \Delta f_D(t)=f_{\rm rx}(t)-f_m
  =
  -\frac{2f_m}{c}\,v_r(t),
  \qquad
  v_r(t)=-\frac{c}{2f_m}\Delta f_D(t).
\label{eq:doppler-vr}
\end{equation}
At \(f_m=1~\mathrm{GHz}\), a line-of-sight velocity of
\(1~\mathrm{km\,s^{-1}}\) gives \(|\Delta f_D|=6.67~\mathrm{kHz}\). The digital lock-in therefore derotates the received phase with a model-assisted local oscillator,
\[
  \Phi_{\rm LO}(t)=2\pi f_m t-\frac{4\pi f_m}{c}R_{\rm model}(t),
\]
and the residual phase and slope are fitted for corrections to
\(R_{\rm model}\) and \(v_r\). A residual derotation error \(\delta v\) gives \(\delta f_D=2f_m\delta v/c\) and must be small enough that coherent-amplitude loss and phase bias remain below the covariance allocation.

Because a $2\pi$ wrap in the envelope phase corresponds, via Eq.~\eqref{eq:phiRF}, to a change $\delta R_{\rm tot} = c/(2 f_m)$ in the one--way path, the estimator~\eqref{eq:Robs} is ambiguous modulo
\begin{equation}
  \Delta R_{\rm amb} = \frac{c}{2 f_m},
  \label{eq:ambwidth}
\end{equation}
(for example, $\Delta R_{\rm amb} = 0.15$\,m at $f_m = 1$\,GHz).  In terms of the underlying round--trip optical path $2R_{\rm tot}$ this corresponds to an ambiguity interval $2\Delta R_{\rm amb} = c/f_m$. 

The photon shot-noise limits for fixed \(f_m\) and lock-in SNR
\({\rm SNR}_{\tt AM}(T)\), where \({\rm SNR}_{\tt AM}(T)\) is the total phase
SNR accumulated over the same integration window \(T\), are
\begin{equation}
  \sigma_{R,{\rm shot}}(T)
  =
  \frac{c}{4\pi f_m}\,
  \frac{1}{{\rm SNR}_{\tt AM}(T)},
  \qquad
  \sigma_{v_r,{\rm shot}}(T)
  =
  \frac{c}{4\pi f_m}\,
  \frac{\sqrt{12}}{T\,{\rm SNR}_{\tt AM}(T)} .
  \label{eq:shot}
\end{equation}
The factor \(\sqrt{12}\) is the standard penalty for estimating a linear slope
from uniformly weighted phase samples over a finite interval. Equivalently, for
\(N=T f_s\) equally spaced samples with per-sample phase variance
\(\sigma_\phi^2\), least-squares regression on
\(\phi_k=a+b t_k+\eta_k\), with the time origin at the center of the fit
window, gives
\begin{equation}
  {\rm Var}\bigl(\widehat b\bigr)
  =
  \frac{12\,\sigma_\phi^2}{N\,T^2}.
  \label{eq:lsq}
\end{equation}
Since \({\rm SNR}_{\tt AM}(T)=\sqrt{N}/\sigma_\phi\) for white phase noise,
Eq.~\eqref{eq:lsq} gives the second expression in Eq.~\eqref{eq:shot}. Thus
\(\sigma_R\propto T^{-1/2}\), while the photon-limited slope estimate scales
as \(T^{-3/2}\).

For a single tone with lock--in  ${\rm SNR}_{\tt AM}$, efficient estimators satisfy ${\rm Var}(\hat\phi)\simeq 1/{\rm SNR}_{\tt AM}^2$ \cite{GoodmanStatOptics,KayEstimation,RifeBoorstyn1974}. For equally spaced samples $\{t_k\}$ over $T$, $\phi_k=a+b t_k+\eta_k$ with ${\rm Var}(\eta_k)=\sigma_\phi^2$ gives ${\rm Var}(b)=12\,\sigma_\phi^2/(N T^2)$, consistent with Eq.~(\ref{eq:lsq}).

\subsection{Oscillator slew across the round trip}
\label{sec:slew}

If the modulation frequency in (\ref{eq:phiRF}) depends on time behaving as $f_m(t) = f_m + \dot f_m t$ during the two--way light time $\tau_{2{\rm w}} \approx 2.56$\,s, then $d\phi/dt$ contains
$2\pi\tau_{2{\rm w}}\dot f_m$, producing a deterministic range--rate bias
\begin{equation}
  \delta v_r \simeq
  \frac{c\,\tau_{2{\rm w}}}{2 f_m}\,\dot f_m.
  \label{eq:slew}
\end{equation}
At $f_m = 1$\,GHz, $|\dot f_m| \lesssim 2.6\times 10^{-6}$\,Hz\,s$^{-1}$ keeps $|\delta v_r| < 10^{-6}$\,m\,s$^{-1}$.
In practice, $f_m(t)$ is recorded and this term is subtracted explicitly; residual oscillator noise is then governed by
the short--term Allan variance of the reference~\cite{Allan1966ProcIEEE}.

Operationally, we record the modulation frequency history $f_m(t)$ and subtract the deterministic contribution of Eq.~(\ref{eq:slew}) in post-processing. The residual stochastic component is then governed by the short-term Allan deviation $\sigma_y(\tau)$ of the maser; with  $\sigma_y(1$--$10~\mathrm{s})\lesssim 10^{-14}$ the corresponding range contribution is bounded by Eq.~(\ref{eq:sigma_R_osc}) at the $\sim$few-$\mu$m level and is absorbed into ${\bf C}_{\rm osc}(T)$.

\subsection{Differential observables}
\label{sec:diffobs}

For two corner--cube reflectors $A$ and $B$ at small angular separation $\theta$, the internal--reference--corrected
phases on a given tone may be written as
\begin{equation}
  \tilde\phi^{(j)}(t_k) =
  \frac{4\pi f_m^\ast}{c}
  \Big(
    R_{\rm geom}^{(j)}(t_k)
    + R_{\rm atm}^{(j)}(t_k)
  \Big)
  + \delta\phi_{\rm nl}^{(j)}(t_k)
  + n_\phi^{(j)}(t_k),
  \qquad j\in\{A,B\},
\end{equation}
where common instrumental terms have been removed by the internal reference and where any residual instrument
nonlinearity is included in $\delta\phi_{\rm nl}^{(j)}$.  The differential phase
\begin{equation}
  \Delta\phi(t_k) \equiv
  \tilde\phi^{(A)}(t_k) - \tilde\phi^{(B)}(t_k)
\end{equation}
then yields the differential two--way range and one--way range--rate over an integration window $T$ via
\begin{equation}
  \Delta \widehat{R}_{2{\rm w}}(T)
  = \frac{c}{4\pi f_m^\ast}\,\overline{\Delta\phi}(T),
  \qquad
  \Delta \widehat{v}_r(T)
  = \frac{c}{4\pi f_m^\ast}\,\widehat{\Delta b}(T),
\end{equation}
where the overbar denotes the mean over $T$ and $\widehat{\Delta b}$ is the slope from a linear regression of $\Delta\phi(t)$ on time over the same window (with $\Delta\widehat{b} = {d\Delta \phi}/{dt}$ being the differential phase slope fit, similar to Eq.~(\ref{eq:vrob})).

The differential two--way range precision can be decomposed as
\begin{equation}
  \sigma_{\Delta R}^2(T) =
  2\,\sigma_{R,{\rm shot}}^2(T)
  + \sigma_{\Delta R,{\rm atm}}^2(T,\theta)
  + \sigma_{\Delta R,{\rm inst}}^2(T,\theta),
  \label{eq:diffbudget}
\end{equation}
with the corresponding expression for the differential range--rate,
\begin{equation}
\sigma^2_{\Delta v_r}(T) = 2\,\sigma^2_{v_r,{\rm shot}}(T) + \sigma^2_{\Delta v_r,{\rm atm}}(T,\theta) + \sigma^2_{\Delta v_r,{\rm inst}}(T,\theta),
  \label{eq:diffbudget-rr}
\end{equation}
where $\sigma_{v_r,{\rm shot}}(T)$ is given by Eq.~(\ref{eq:shot}) and the atmospheric and instrumental terms inherit their $T$ and $\theta$ dependence from the same turbulence and metrology models used for
$\sigma^2_{\Delta R,{\rm atm}}$ and $\sigma^2_{\Delta R,{\rm inst}}$.

The first term in Eq.~\eqref{eq:diffbudget} imposes a hard lower bound. For statistically independent photon noise in the two reflector returns,
\begin{equation}
  \sigma_{\Delta R,{\rm shot}}
  =
  \left[
    \sigma^2_{R,{\rm shot},A}
    +
    \sigma^2_{R,{\rm shot},B}
  \right]^{1/2},
\label{eq:diff-photon-floor}
\end{equation}
which reduces to \(\sqrt{2}\,\sigma_{R,{\rm shot}}\) for equal links. Hence
\(\sigma_{\Delta R}\ge\sigma_{\Delta R,{\rm shot}}\) regardless of atmospheric
or instrumental common-mode rejection. At \(f_m=1~\mathrm{GHz}\),
\(T=100~\mathrm{s}\), and negligible background, the equal-link differential
photon floors for Cases A--C are \(115\)--\(191~\mu\mathrm{m}\),
\(43\)--\(78~\mu\mathrm{m}\), and \(31\)--\(43~\mu\mathrm{m}\), respectively.
A \(20~\mu\mathrm{m}\) differential target requires, before adding atmosphere
or instrument terms,
\begin{equation}
  \dot N_\gamma
  \gtrsim
  \frac{1}{T}
  \left[
    \frac{2\sqrt{2}}{a_m}
    \frac{c}{4\pi f_m}
    \frac{1}{\sigma_{\Delta R}}
  \right]^2
  =
  2.3\times10^5~\mathrm{s^{-1}}
  \left(\frac{0.7}{a_m}\right)^2
  \left(\frac{100~\mathrm{s}}{T}\right)
  \left(\frac{20~\mu\mathrm{m}}{\sigma_{\Delta R}}\right)^2 .
\label{eq:diff-20um-flux}
\end{equation}
Thus \(20~\mu\mathrm{m}\) differential range is a stretch requirement, not a nominal Case~B result.

For Kolmogorov turbulence, after each reflector's phase has been averaged over an analysis window \(T\), a useful parametrization of the residual differential atmospheric term is
\begin{equation}
  \sigma^2_{\Delta R,\rm atm}(T,\theta)
  \simeq
  K_R^2
  \left(\frac{\theta}{\theta_0}\right)^{5/3}
  \frac{\tau_0}{T},
\label{eq:diffatm}
\end{equation}
where \(K_R\) is a site-dependent amplitude, \(\tau_0\) is the short-time turbulence correlation time that sets the statistical averaging law, and \(\theta_0\) is an effective angular decorrelation scale for differential path delay. The parameter \(\theta_0\) is not identified with the adaptive-optics wavefront isoplanatic angle; it is an empirical path-delay scale to be
calibrated from site telemetry, LLR/SLR residuals, or dedicated A/B reflector tests. Eq.~\eqref{eq:diffatm} does not require the A/B switching cadence to be shorter than the millisecond-scale optical turbulence time. Fast turbulence is treated statistically through the factor \(\tau_0/T\), while the A/B cadence must be short compared with slow differential-delay drift and must preserve correct transmit/receive assignment over the
\(\tau_{2\mathrm{w}}\simeq2.56~\mathrm{s}\) lunar round trip. For \(\theta\lesssim0.1^\circ\) and \(T\sim10^2~\mathrm{s}\), the calibrated differential atmospheric contribution can be in the \(\sim10\)--\(50~\mu \mathrm{m}\) band under good-to-excellent conditions, consistent with the differential budgets in Sec.~\ref{sec:perf}.

Note that, in terms of the joint covariance matrix $\Cov(T)$ from Eq.~\eqref{eq:covsplit}, the scalar range budget in Eq.~\eqref{eq:budget}, for instance, can be written as
\begin{equation}
  \sigma_R^2(T) = C_{11}(T)
  = C_{\mathrm{shot},11}(T) + C_{\mathrm{atm},11}(T) + C_{\mathrm{inst},11}(T),
\end{equation}
with the oscillator and nonlinearity contributions ${\bf C}_{\mathrm{osc}}(T)$ and ${\bf C}_{\mathrm{nl}}(T)$ either folded into $\sigma_{R,{\rm inst}}^2(T)$ or treated explicitly when needed.  For the differential case, the same structure applies but with the atmospheric and instrumental terms replaced by their differential counterparts $\sigma_{\Delta R,{\rm atm}}^2(T,\theta)$ and $\sigma_{\Delta R,{\rm inst}}^2(T,\theta)$ as in
Eqs.~(\ref{eq:diffbudget})--(\ref{eq:diffatm}).

\section{Ambiguity removal: multi-tone synthetic wavelengths}
\label{sec:unwrap}

The envelope phase $\phi(t)$ in Eq.~\eqref{eq:phiRF} is defined modulo
$2\pi$, which maps to a range ambiguity in the estimator
$\widehat{R}_{2\mathrm{w}}$ of
\begin{equation}
  \Delta R_{2\mathrm{w}} = \frac{c}{2 f_m},
  \label{eq:rhoambiguity}
\end{equation}
i.e.\ a change of $2\pi$ in the measured phase corresponds to a change $\Delta R_{2\mathrm{w}}$ in the inferred one--way geometric range and to a change $2\Delta  R_{2\mathrm{w}} = c/f_m$ in the modeled round--trip
optical path $2R_{\rm tot}$.

With tones $\{f_i\}$, ambiguity is lifted using synthetic wavelengths
\begin{equation}
  \Lambda_{2{\rm w},ij} =
  \frac{c}{2|f_i-f_j|},
  \qquad
  \Lambda_{1{\rm w},ij} =
  \frac{c}{4|f_i-f_j|}.
  \label{eq:syntw}
\end{equation}

An illustrative tone plan and the associated ambiguity intervals and synthetic wavelengths for the tone set used in this work are summarized in Table~\ref{tab:tones-synth}.

\begin{table}[t]
  \centering
  \caption{Example modulation-tone set and associated one-way ambiguity intervals
  $\Delta R_{\rm amb}=c/(2f_m)$ and two-way synthetic wavelengths
  $\Lambda_{2{\rm w},ij}=c/(2|f_i-f_j|)$, as defined in Eqs.~\eqref{eq:rhoambiguity}
  and~\eqref{eq:syntw}.  Numerical values assume
  $c = 299\,792\,458~\mathrm{m\,s^{-1}}$ and are rounded to three significant digits.
  The close pair at 50/50.1~MHz provides a kilometre-scale synthetic wavelength that
  greatly simplifies integer ambiguity resolution.}
  \label{tab:tones-synth}
  \begin{tabular}{lccc}
    \hline
    Tone / pair &
    Frequency & 
    $\Delta R_{\rm amb} = c/(2 f_m)$ [m] &
    $\Lambda_{2{\rm w}} = c/\bigl(2|f_i-f_j|\bigr)$ [m] \\
    \hline\hline
    Low tone &
    $f_{\rm L} = 50~\mathrm{MHz}$ &
    $2.998$ & -- \\
    Intermediate tone &
    $f_{\rm I} = 200~\mathrm{MHz}$ &
    $0.749$ & -- \\
    Fine (precision) tone &
    $f_{\rm H} = 1\,000~\mathrm{MHz}$ &
    $0.150$ & -- \\
    Synthetic (L,\,I) &
    $|f_{\rm I}-f_{\rm L}| = 150~\mathrm{MHz}$ &
    -- & $0.999$ \\
    Synthetic (L,\,H) &
    $|f_{\rm H}-f_{\rm L}| = 950~\mathrm{MHz}$ &
    -- & $0.158$ \\
    Close synthetic pair &
    $|50.1-50.0| = 0.1~\mathrm{MHz}$ &
    -- & $1.50\times 10^{3}$ \\
    \hline
  \end{tabular}
\end{table}

A practical implementation uses a small set of tones, for example \(\{50, 50.1, 200, 1000\}\,\mathrm{MHz}\), and resolves the integer ambiguity via a constrained search over the synthetic-wavelength bins. Operationally, the procedure can be summarized as follows (it is analogous to multi--frequency ranging in SLR and RF navigation~\cite{Degnan1993ProcIEEE}):
\begin{enumerate}
  \item Use the lowest-frequency tone to obtain a coarse estimate of the unwrapped phase and hence of the two-way range modulo the corresponding ambiguity interval. For a tone near \(50~\mathrm{MHz}\) this interval is of order a few metres.
  \item Form a synthetic-wavelength observable from the close frequency pair at \(50\) and \(50.1~\mathrm{MHz}\). The associated synthetic two-way wavelength \(\Lambda_{2\mathrm{w}}\) is of order kilometres, and the measured phase difference between the two tones constrains the admissible set of integer ambiguities for the coarse range solution.
  \item Use the intermediate-frequency tone (e.g. \(200~\mathrm{MHz}\)) to refine the range estimate within the remaining synthetic-wavelength bins. The shorter ambiguity interval at this frequency restricts the allowed integer combinations further, still subject to consistency with the dynamical light-time model.
  \item Finally, use the highest-frequency tone (e.g. \(1~\mathrm{GHz}\)) as the precision carrier. The admissible integer for this tone is selected such that the corresponding range solution is simultaneously consistent with the coarse and intermediate-frequency constraints and with the predicted round-trip light time.
  \item Reject any integer combination for which the implied range differs from the modelled light time by more than the allocated synthetic-wavelength tolerance or for which the multi-tone residuals indicate unmodelled frequency-dependent path delays.
\end{enumerate}
In this scheme the internal metrology and calibration keep the frequency-dependent instrumental path differences well below the synthetic-wavelength scale, so that the multi-tone integer search is dominated by the photon noise and the dynamical model rather than by hardware nonlinearity.

Frequency-dependent RF and optical delays affect two distinct requirements.
First, for integer ambiguity resolution, the calibrated inter-tone path
difference must be small compared with the relevant synthetic wavelength,
\begin{equation}
  \left|\Delta R_{\rm inst}(f_i,f_j)\right|
  \lesssim
  \eta_\Lambda\,\Lambda_{2{\rm w},ij},
  \qquad
  \eta_\Lambda\simeq 0.05\mbox{--}0.1 .
\label{eq:ambiguity-nl}
\end{equation}
This centimetre-to-kilometre scale condition prevents selection of the wrong
synthetic-wavelength bin. It is not a micrometer-level precision requirement.
Second, after the correct integer has been selected, the residual calibrated
phase error on the precision tone \(f_H\) must satisfy
\begin{equation}
  \sigma_{R,{\rm nl}}(T)
  =
  \frac{c}{4\pi f_H}\,
  \sigma_{\phi,{\rm nl}}(T)
  =
  \sqrt{C_{{\rm nl},11}(T)}
  \lesssim 20\mbox{--}25~\mu\mathrm{m}.
\label{eq:sig_r}
\end{equation}
At \(f_H=1~\mathrm{GHz}\), this corresponds to
\[
  \sigma_{\phi,{\rm nl}}(T)
  \lesssim
  1.0\times 10^{-3}~\mathrm{rad}
  \left(\frac{\sigma_{R,{\rm nl}}}{25~\mu\mathrm{m}}\right).
\]
Thus Eq.~\eqref{eq:ambiguity-nl} is the integer-bin requirement, while Eq.~\eqref{eq:sig_r} is the precision-tone calibration requirement entering \({\bf C}_{\rm nl}(T)\).

\section{Photon--counting lock-In SNR and link budget}
\label{sec:snr}

Let post--filter signal and background be \(\dot N_\gamma\) and \(\dot N_b\) (both in s\(^{-1}\)), with AM depth \(a_m\) (the same amplitude--modulation index introduced in (\ref{eq:transm})).  Throughout this section we follow the notation of the high--power CW link study of Ref.~\cite{Turyshev:CW-LLR:2025}: $\dot N_\gamma$ and $\dot N_b$ denote the detected signal and noise photon rates at the output of the receive chain, and $\eta_{\rm eff}$ is the same end--to--end efficiency parameter used there, combining telescope throughput, atmospheric transmission, CCR response, stellar--aberration loss, and detector quantum efficiency into a single scalar factor.

A sinusoidal lock--in at $f_m$ accumulates a lock--in signal-to--noise ratio ${\rm SNR}_{\tt AM}(T)$:
{}
\begin{equation}
{\rm SNR}_{\tt AM}(T) \simeq \tfrac12 a_m
  \sqrt{\frac{\dot N_\gamma T}{1 + \dot N_b / \dot N_\gamma}}=\tfrac12 a_m
{\rm SNR}_{\rm 1\,s}  \sqrt{\frac{T}{\rm 1\,s}},
  \label{eq:lockin_snr}
\end{equation}
consistent with standard phasor--sum statistics for Poisson processes~\cite{GoodmanStatOptics,KayEstimation,RifeBoorstyn1974}.  In the notation of Ref.~\cite{Turyshev:CW-LLR:2025}, ${\rm SNR}_{\tt AM}(T)$ plays the role of the total SNR over an integration window~$T$, with ${\rm SNR}_{1\,{\rm s}}=  [{\dot N_\gamma}/(1 + \dot N_b / \dot N_\gamma)]^\frac{1}{2}$ (see (29) in \cite{Turyshev:CW-LLR:2025}) corresponding approximately to ${\rm SNR}_{\tt AM}(1\,{\rm s})$ once the factor $\tfrac12 a_m$ associated with the AM depth in Eq.~\eqref{eq:lockin_snr} is taken into account. At $f_m = 1$\,GHz, $c/(4\pi f_m) = 2.38567\times 10^{-2}$\,m\,rad$^{-1}$; thus $\sigma_{R_{2{\rm w}}} = 0.10$\,mm requires ${\rm SNR}_{\tt AM}(100\,{\rm s}) \approx 2.4\times 10^2$.

In deriving Eq.~\eqref{eq:lockin_snr} we treat the detected photon stream as a stationary Poisson process over the window \(T\). Atmospheric scintillation then enters primarily through slow fluctuations of \(\dot N_\gamma\) and hence of the measured \({\rm SNR}_{\tt AM}(T)\); windows with deep fades or
background excursions are rejected by the quality gates.

The same phase estimator can be implemented with either time-tagged photon counting or a linear RF receiver. In photon-counting mode the detector does not produce an analog 1~GHz photocurrent waveform. Instead, the arrival times
\(t_n\) are time tagged relative to the maser-referenced modulation phase and the per-tone phasor is formed digitally.

The phasor is formed after model-assisted derotation of the expected AM-envelope
Doppler. Thus the quantity accumulated in a window is
\begin{equation}
  Z_m(T)=\sum_{n=1}^{N_{\rm det}}
  \exp[-i\,\Phi_{\rm LO}(t_n)],
\label{eq:photon-phasor}
\end{equation}
where \(\Phi_{\rm LO}(t)\) is the same model phase used in
Eqs.~\eqref{eq:am-doppler}--\eqref{eq:doppler-vr}. In the derotated frame the residual photon rate may be written as
\[
  \lambda(t)=\dot N_\gamma[1+a_m\cos(\delta\phi(t))]+\dot N_b ,
\]
where \(\delta\phi(t)\) is the slowly varying residual phase to be estimated. For stationary rates and small residual bandwidth,
\[
  |\langle Z_m\rangle|=(a_m/2)\dot N_\gamma T,
  \qquad
  {\rm Var}(Z_m)\simeq(\dot N_\gamma+\dot N_b)T,
\]
which gives Eq.~\eqref{eq:lockin_snr}. Without this derotation, a lunar line-of-sight velocity of \(1~\mathrm{km\,s^{-1}}\) would shift a \(1~\mathrm{GHz}\) envelope by \(6.67~\mathrm{kHz}\), preventing coherent accumulation on long windows. Finite single-event timing jitter \(\sigma_t\) reduces the effective
modulation depth by
\begin{equation}
  a_{m,\rm eff}
  =
  a_m\exp\!\left[-\frac{1}{2}(2\pi f_m\sigma_t)^2\right].
\label{eq:jitter-contrast}
\end{equation}
At \(f_m=1~\mathrm{GHz}\), \(\sigma_t=50~\mathrm{ps}\) gives
\(a_{m,\rm eff}/a_m=0.952\), while \(100~\mathrm{ps}\) gives \(0.821\).

For a linear InGaAs or optically assisted receiver, analog bandwidth through the highest modulation tone is required. The corresponding RF-band electronics noise should satisfy
\begin{equation}
  {\rm NEP}_{\rm el}
  \lesssim
  {\rm NEP}_{\rm ph}
  =
  \left(2E_\gamma P_1\right)^{1/2},
  \qquad
  P_1=\dot N_\gamma E_\gamma,\quad
  E_\gamma=\frac{hc}{\lambda}.
\label{eq:linear-nep}
\end{equation}
At 1064~nm, \(E_\gamma=1.87\times10^{-19}~\mathrm{J}\). For Case~B,
\(\dot N_\gamma=(3\)--\(5)\times10^4~\mathrm{s^{-1}}\), so
\(P_1=5.6\)--\(9.3~\mathrm{fW}\) and
\({\rm NEP}_{\rm ph}=4.6\)--\(5.9\times10^{-17}~\mathrm{W\,Hz^{-1/2}}\).

The minimum photon-limited integration time for a target absolute range precision \(\sigma_R\) is
\begin{equation}
  T_{\rm min}(\sigma_R)
  =
  \Big[
    \frac{2}{a_m\sqrt{\dot N_\gamma}}\,
    \frac{c}{4\pi f_m\sigma_R}
  \Big]^2
  \Big(1+\frac{\dot N_b}{\dot N_\gamma}\Big).
\label{eq:Tmin}
\end{equation}
For negligible background, \(f_m=1~\mathrm{GHz}\), and \(a_m=0.5\)--0.7, \(T_{\rm min}(0.10~\mathrm{mm})\) is \(66\)--\(182~\mathrm{s}\) in Case~A, \(9\)--\(30~\mathrm{s}\) in Case~B, and \(5\)--\(9~\mathrm{s}\) in Case~C.
For \(30~\mu\mathrm{m}\), the corresponding times are
\(0.73\)--\(2.0~\mathrm{ks}\), \(0.10\)--\(0.34~\mathrm{ks}\), and
\(52\)--\(101~\mathrm{s}\).

Higher detected photon rate provides operational flexibility, but it should not be interpreted as making the stochastic atmospheric average smaller on a shorter window. In the Kolmogorov model of Appendix~\ref{app:atm_turbulence}, \(\sigma^2_{R,\rm atm}(T)\propto \tau_0/T\) for \(T\gg\tau_0\); longer averages reduce the white-in-time atmospheric contribution. Shorter windows are
useful only when they reduce exposure to slow nonstationary drifts, cycle-slip risk, or changing observing conditions.

Combining the beam geometry, two-way transmission, and collection aperture yields an approximate flux at the
detector may given as below
\begin{equation}
  \dot N_\gamma \simeq
  \frac{P_0}{E_\gamma}\,\eta_{\rm eff}\,
  \frac{A_{\rm CCR}}{A_{\rm spot,Moon}}\,
  \frac{A_{\rm tel}}{A_{\rm spot,Earth}},
  \qquad
  E_\gamma = \frac{hc}{\lambda},
  \label{eq:link}
\end{equation}
where $P_0$ is the transmitter power, $E_\gamma$ is the photon energy, $\eta_{\rm eff}$ is the end--to--end efficiency, and $A_{\rm spot,Moon}$ and $A_{\rm spot,Earth}$ are the footprint areas set by diffraction and turbulence~\cite{Degnan1993ProcIEEE,Turyshev:CW-LLR:2025}. 

Consistent with the AM--CW power model in Sec.~\ref{sec:model}, the mean received optical power $P_1$ in \eqref{eq:Prx} can be identified with the power associated with the detected photon rate in \eqref{eq:link} via $P_1 \simeq \dot N_\gamma E_\gamma$, with the detector quantum efficiency and other throughput factors already absorbed into $\eta_{\rm eff}$ in the same way as in \cite{Turyshev:CW-LLR:2025}.  This makes explicit that the modulation envelope in \eqref{eq:Prx} is driven by the same link budget that sets the photon--statistical term $\Cshot(T)$ in \eqref{eq:covsplit}.

\begin{table}[h]
\centering
\caption{Representative link parameters used in the photon--flux estimates based on Eq.~\eqref{eq:link}.  Values correspond to the 1~kW, 1064~nm, 1~m-aperture, 10~cm CCR baseline adopted from Ref.~\cite{Turyshev:CW-LLR:2025}.}
\label{tab:link-budget}
\begin{tabular}{lcl}
\hline
Quantity & Symbol & Representative value \\
\hline\hline
Transmitter power & $P_0$ & $1~\mathrm{kW}$ \\
Optical wavelength & $\lambda$ & $1064~\mathrm{nm}$ \\
Telescope aperture & $D$ & $1~\mathrm{m}$ \\
CCR diameter & $d_{\rm CCR}$ & $10~\mathrm{cm}$ \\
End-to-end efficiency & $\eta_{\rm eff}$ & $0.2$--$0.4$ \\
Round-trip light time & $\tau$ & $\simeq 2.56~\mathrm{s}$ \\
Signal photon rate (design range) & $\dot N_\gamma$ & $\sim 10^{3}$--$10^{5}~\mathrm{s^{-1}}$ \\
\hline
\end{tabular}
\end{table}

For the 1~kW, 1064~nm, 1~m-aperture, 10~cm CCR baseline of Ref.~\cite{Turyshev:CW-LLR:2025}, the link calculation corresponding to Eq.~(\ref{eq:link}) yields a detected photon rate
\(
\dot N_\gamma \simeq (5\text{--}7)\times 10^{3}\ {\rm s^{-1}}
\)
for an end-to-end efficiency $\eta_{\rm eff}\simeq 0.2$ and good ($r_0\simeq 0.2$~m) seeing (see Table~\ref{tab:link-budget} for the values used). Inserted into Eq.~(\ref{eq:lockin_snr}), this baseline gives ${\rm SNR}_{\tt AM}(100~{\rm s})\simeq 200$ and a photon-limited two-way precision $\sigma_{R,{\rm shot}}\simeq 0.1$~mm at $f_m=1$~GHz.

\paragraph*{Representative photon--flux regimes.}

To connect the lock--in SNR scalings above to realistic operations, it is useful to group the detected photon rate $\dot N_\gamma$ into three representative regimes that will be used throughout the remainder of the paper. Case~A represents a ``generic'' 1~kW CW station with $\dot N_\gamma\simeq (5$--$7)\times10^{3}$~s$^{-1}$, matching the link budget of the earlier CW study for a 1~m telescope ranging to a single 10~cm CCR under good seeing~\cite{Turyshev:CW-LLR:2025}. Case~B is the dedicated AM--CW configuration adopted for the design allocations below, with modestly larger collecting area and/or higher end--to--end efficiency such that $\dot N_\gamma\sim (3$--$5)\times10^{4}$~s$^{-1}$ under similar seeing. Case~C denotes a photon--rich regime with $\dot N_\gamma\sim 10^{5}$~s$^{-1}$, corresponding to larger apertures and/or improved throughput at a very good site.

\begin{figure}[t]
  \centering
  \includegraphics[width=0.55\columnwidth]{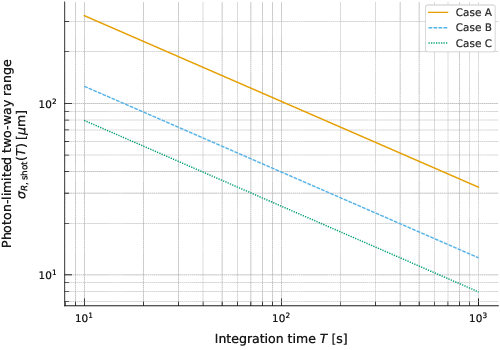}
  \caption{Photon-limited two-way range uncertainty
  $\sigma_{R,\mathrm{shot}}(T)$, from (\ref{eq:shot}) and (\ref{eq:lockin_snr}) with the values from Table~\ref{tab:link-budget},  as a function of integration time $T$ for the three representative photon-flux regimes (Cases~A--C) used in the link and covariance analysis. The curves assume a single GHz-class modulation tone on a bright CW carrier with negligible background, so that all three cases follow the expected $\sigma_{R,\mathrm{shot}}\propto T^{-1/2}$ scaling; the vertical separation reflects the different detected photon rates in each regime.}
  \label{fig:sigmaR-shot-cases}
\end{figure}

Inserting these fluxes into Eq.~\eqref{eq:lockin_snr} gives, for $T\simeq 100$~s, lock--in SNRs of order ${\rm SNR}_{\tt AM}\sim 2\times10^{2}$, $(4$--$8)\times10^{2}$, and $\gtrsim 10^{3}$ for Cases~A, B, and C, respectively. At $f_m=1$~GHz this translates, via Eq.~\eqref{eq:shot}, into photon--limited two--way precisions of $\sigma_{R,{\rm shot}}\simeq 0.1$~mm for Case~A, $\sigma_{R,{\rm shot}}\simeq (3$--$6)\times10^{-5}$~m for Case~B, and $\sigma_{R,{\rm shot}}\lesssim 3\times10^{-5}$~m for Case~C (see Fig.~\ref{fig:sigmaR-shot-cases}). These three flux regimes bracket the range from generic high--power CW LLR to the more ambitious AM--CW station considered here; in Secs.~\ref{sec:atmo} and~\ref{sec:instrum-er} they are paired with the atmospheric and instrumental allocations to form the combined operating Cases~A--C summarized in Table~\ref{tab:three-cases}.

In all three regimes we hold the lunar retroreflector design fixed to a next--generation $10\,$cm CCR geometry, consistent with the earlier high--power CW LLR study, so that reflector--induced pulse spreading, libration smearing, and thermoelastic deformations contribute at or below the atmospheric and instrumental allocations and can be absorbed into the geometric model $R_{\mathrm{geom}}(t;\boldsymbol{\theta})$ rather than treated as additional stochastic terms in $C_{ij}(T)$.

In the design studies below we are interested in a more “photon-rich’’ configuration, e.g. a modestly larger receive aperture and/or improved throughput ($\eta_{\rm eff}\simeq 0.35$--0.4), for which Eq.~(\ref{eq:link}) yields \( \dot N_\gamma \sim (3\text{--}5)\times 10^{4}\ {\rm s^{-1}} \) under good seeing. These higher rates are consistent with the upper end of the link-budget range in \cite{Turyshev:CW-LLR:2025} when scaled to larger collecting area and improved efficiencies (Table~\ref{tab:link-budget}), and they are used here only to set the {\em design} shot-noise allocations (30~$\mu$m for $R_{2w}$ in Table~\ref{tab:summary-2w}), not as a universal value for all 1~m / 10~cm implementations. With $a_m\simeq 0.5$--0.7 and negligible background ($\dot N_b\ll\dot N_\gamma$), Eq.~(\ref{eq:lockin_snr}) then gives
\({\rm SNR}_{\tt AM}(100~{\rm s})\simeq (4\text{--}8)\times 10^{2},\) corresponding to a photon-limited two-way range precision $\sigma_{R,{\rm shot}}\simeq (3\text{--}6)\times 10^{-5}$~m at $f_m=1$~GHz. For somewhat higher return rates, $\dot N_\gamma\sim 10^5\,\mathrm{s^{-1}}$, the lock--in SNR on 100\,s windows exceeds $10^3$ and $\sigma_{R,{\rm shot}}$ drops below $3\times 10^{-5}\,\mathrm{m}$. 

Eq.~\eqref{eq:lockin_snr} treats the detected photon stream as a stationary Poisson process with constant mean rate $\dot N_\gamma$; atmospheric scintillation then enters only through slow fluctuations of $\dot N_\gamma$ and, hence, of ${\rm SNR}_{\tt AM}(T)$.  For a 1\,m aperture at 1064\,nm on an Earth--Moon path the expected scintillation index is modest, so that amplitude noise at the modulation frequency is small compared to photon shot noise on the integration times $T\gtrsim 10$\,s considered here.  In practice we monitor ${\rm SNR}_{\tt AM}(T)$ in real time and reject analysis windows in which deep scintillation fades or background excursions drive the SNR below the thresholds in Table~\ref{tab:gates}.  A more complete treatment of scintillation as multiplicative noise in the photon stream, and its propagation into $\Cshot(T)$, can be incorporated in future work but does not change the design-level allocations adopted in Sec.~\ref{sec:errors}. 

\section{Atmosphere and instrument: precision budgets}
\label{sec:errors}

Two--way precision over $T$ is
\begin{equation}
  \sigma_R^2(T) =
  \sigma_{R,{\rm shot}}^2(T) +
  \sigma_{R,{\rm atm}}^2(T) +
  \sigma_{R,{\rm inst}}^2(T).
  \label{eq:budget}
\end{equation}

In terms of the covariance decomposition in Eq.~\eqref{eq:covsplit}, the scalar precision can be written as
\[
\sigma_R^2(T) = C_{11}(T)
 = C_{{\rm shot},11}(T) + C_{{\rm atm},11}(T) + C_{{\rm inst},11}(T),
\]
with the understanding that the oscillator and nonlinearity contributions are either treated explicitly as $\Cosc(T)$ and $\Cnl(T)$ or, in the regime of interest here, safely absorbed into the instrumental term. For the integration windows $T \simeq 30$--$100$~s that drive most science applications, a representative allocation is
\[
\sigma_{R,{\rm shot}} \simeq 30~\mu{\rm m}, \quad
\sigma_{R,{\rm atm,res}} \simeq 60~\mu{\rm m}, \quad
\sigma_{R,{\rm inst,res}} \simeq 40~\mu{\rm m},
\]
with oscillator and nonlinearity contributions bounded at the few-$\mu$m and tens-of-$\mu$m levels, respectively. These numbers define quantitative targets for the hardware and CONOPS and are used directly in the implementation gates in Sec.~\ref{sec:gates} and in the summary budget in Table~\ref{tab:summary-2w}.

For example, the photon--limited covariance matrix of the joint estimator $(R_{2{\rm w}},v_r)$ over a window
$T$ may be written 
\begin{equation}
  \mathbf{C}_{\rm shot}(T) =
  \begin{pmatrix}
    \sigma_{R,{\rm shot}}^2(T) &
    {\rm cov}\bigl(R_{2{\rm w}},v_r;T\bigr) \\
    {\rm cov}\bigl(R_{2{\rm w}},v_r;T\bigr) &
    \sigma_{v_r,{\rm shot}}^2(T)
  \end{pmatrix},
  \label{eq:covshot}
\end{equation}
where $\sigma_{R,{\rm shot}}(T)$ and $\sigma_{v_r,{\rm shot}}(T)$ are given by (\ref{eq:shot}).  For equally spaced samples and a linear regression model of (\ref{eq:lsq}), the off--diagonal covariance can be made negligible by choosing the time origin at the center of the fit window, so that $\sum_k t_k = 0$.  In that case the range and range--rate estimates are effectively uncorrelated at the photon--noise level.

\subsection{Representative error budget}

\subsubsection{Atmosphere}
\label{sec:atmo}

The relevant quantity for LLR is the variation of the neutral-atmosphere delay over an integration window $T$, not its absolute value. Under Kolmogorov turbulence with frozen flow, the optical-path structure function obeys the usual $(\tau/\tau_0)^{5/3}$ scaling, and for $T\gg \tau_0$ the variance of the time-averaged path scales as $\sigma_{R,\mathrm{atm}}^2(T)\propto \tau_0/T$, as summarized in the atmospheric-turbulence Appendix~\ref{app:atm_turbulence}. Calibrating the overall amplitude of this model to existing mm-class LLR data and SLR experience (coherence time \(\tau_0\sim 5~\mathrm{ms}\), effective differential-delay angular scale \(\theta_0\sim 1^\circ\), and Fried parameter \(r_0\simeq0.2~\mathrm{m}\) at 1064 nm) gives
\[
  \sigma_{R,\mathrm{atm}}(T)\sim (3\text{--}5)\times10^{-4}~\mathrm{m}
\]
on $T\simeq 30$--$100$~s for typical mid-latitude seeing without aggressive elevation cuts. This reproduces the $300$--$500~\mu\mathrm{m}$ atmospheric term that dominates the generic 1~kW CW LLR error budget in \cite{Turyshev:CW-LLR:2025} and corresponds to our Case~A.

These scalings are essentially independent of the AM--CW architecture itself; they are set by the site, elevation, and weather cuts. Higher detected photon rate does not make the stochastic atmospheric average smaller on a shorter window. In the Kolmogorov model used here, \(\sigma^2_{R,\mathrm{atm}}(T)\propto\tau_0/T\) for \(T\gg\tau_0\), so the
white-in-time contribution decreases with longer averaging. The role of the higher-flux AM--CW link is instead to provide operational flexibility: the station can use longer windows while maintaining photon SNR, thereby averaging the stochastic atmosphere down, or it can use shorter windows when necessary to reduce exposure to slow nonstationary drifts, changing seeing, or cycle-slip risk. The Case~B and Case~C atmospheric allocations therefore rely on site selection, elevation and SNR cuts, turbulence monitoring, and internal metrology, not on a claim that shorter coherent windows by themselves reduce the Kolmogorov variance.

\begin{figure}[t]
  \centering
  \includegraphics[width=0.50\columnwidth]{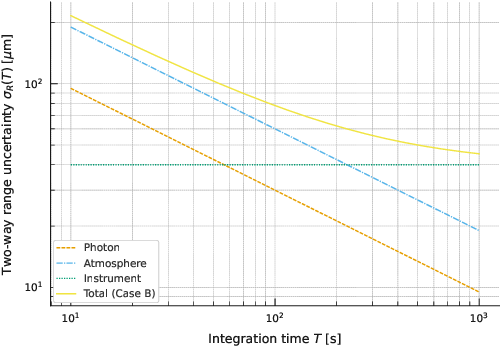}
  \caption{Representative absolute two-way range error budget $\sigma_R(T)$ for the dedicated AM--CW facility (Case~B), from (\ref{eq:budget}) and the values from Table~\ref{tab:link-budget}. The photon, atmosphere, and instrument contributions are shown separately together with their root-sum-square total. Photon
and atmospheric terms follow the $T^{-1/2}$ scaling implied by the shot-noise and Kolmogorov models, while the instrumental term is approximately constant over the $30$--$100~\mathrm{s}$ band. The total curve illustrates the transition from a photon/atmosphere-dominated regime at short $T$ to an instrumental floor at longer integrations, consistent with the scalar allocations used in the range covariance.}
  \label{fig:range-budget-caseB}
\end{figure}

With site selection, elevation $e\gtrsim 30^\circ$, and explicit SNR and quality cuts, the same model yields much smaller fluctuations:
\[
  \sigma_{R,\mathrm{atm}}(T)\sim (5\text{--}15)\times10^{-5}~\mathrm{m},
\]
i.e. $\sim 50$--$150~\mu\mathrm{m}$ on $T=10$--$100$~s. These values are consistent with APOLLO and near-infrared LLR experience and underlie the $\sim 60~\mu\mathrm{m}$ atmospheric allocation adopted for the dedicated AM--CW facility (Case~B, see Fig.~\ref{fig:range-budget-caseB}). 

For the most favorable conditions at a very good site (top quartile of the seeing distribution, somewhat larger $r_0$ than assumed above), the same scalings imply that the residual atmospheric contribution can realistically be driven into the $\sim 30$--$80~\mu\mathrm{m}$ band on $T\simeq 30$--$100$~s, which we associate with Case~C.

Calibrating the Kolmogorov model for the range-averaged path fluctuations to existing mm-class LLR and SLR experience is most transparent when expressed in terms of a small set of seeing parameters. Table~\ref{tab:atm-calibration} summarizes representative combinations of Fried parameter, coherence time, and effective differential-delay angular scale together with the corresponding ranges of absolute and differential atmospheric residuals on the analysis windows used throughout the paper. The angular scale \(\theta_0\) is an empirical decorrelation scale for differential path delay, not the adaptive-optics wavefront isoplanatic angle.

By construction, Regimes~A, B, and C in Table~\ref{tab:atm-calibration} are paired with the photon--flux regimes of Sec.~\ref{sec:snr} to define the atmospheric contributions to operating Cases~A, B, and C in Table~\ref{tab:three-cases}.

\begin{table}[t]
  \centering
  \caption{Illustrative calibration of the Kolmogorov atmospheric model to representative observing conditions. Each regime specifies a Fried parameter $r_0$ at 1064~nm, coherence time $\tau_0$, and effective differential-delay angular scale $\theta_0$ together with the resulting ranges for the absolute and differential two--way range residuals on $T$ in the tens--of--seconds regime. Regimes~A, B, and C provide the atmospheric components of operating Cases~A, B, and C in Table~\ref{tab:three-cases}, respectively.}
  \label{tab:atm-calibration}
  \begin{tabular}{lcccc}
    \hline
    Regime & \(r_0\) at 1064 nm & \(\tau_0\) & \(\theta_0\) & Typical residuals \\
    \hline\hline
 A:   Generic mid-latitude & \(\sim 0.2~\mathrm{m}\) & \(\sim 5~\mathrm{ms}\) & \(\sim 1^\circ\) &
    \(\sigma_{R,\mathrm{atm}}(T) \sim 300\text{--}500~\mu\mathrm{m}\) \\
B:    Dedicated AM--CW site & \(\gtrsim 0.2~\mathrm{m}\) & few\(\times 10~\mathrm{ms}\) & \(\sim 1^\circ\) &
    \(\sigma_{R,\mathrm{atm}}(T) \sim 50\text{--}150~\mu\mathrm{m}\) \\
C:    Excellent seeing & \(\gtrsim 0.3~\mathrm{m}\) & \(\gtrsim 10~\mathrm{ms}\) & \(\gtrsim 1^\circ\) &
    \(\sigma_{R,\mathrm{atm}}(T) \sim 30\text{--}80~\mu\mathrm{m}\) \\
    \hline
    & & & &
    \(\sigma_{\Delta R,\mathrm{atm}}(T,\theta) \sim 10\text{--}50~\mu\mathrm{m}\) for \(\theta \lesssim 0.1^\circ\) \\
    \hline
  \end{tabular}
\end{table}

Although the error budget in Eq.~\eqref{eq:budget} and Table~\ref{tab:summary-2w} assumes single-wavelength operation at 1064\,nm, the AM--CW architecture is compatible with dual-wavelength (or ``two-color'') ranging as an optional refinement. In that case a second, nearby optical wavelength---either from a separate seed or from frequency conversion of the primary source---would be transmitted through the same telescope and processed through the same RF/ADC chain, providing an additional set of envelope phases and slopes. Because the neutral atmosphere is only weakly dispersive in the near infrared, dual-wavelength operation does not eliminate the tropospheric delay, but the small, well-modeled chromatic dependence of the refractive index can be exploited to constrain residual errors in $R_{\rm atm}(t)$ and in the mapping functions~\cite{Degnan1993ProcIEEE}. In the design presented here, the allocations for $\sigma_{R,{\rm atm}}(T)$ can be met with site selection, elevation cuts, and modern mapping functions alone; a dual-wavelength option would primarily provide redundancy and a path to tightening atmospheric systematics in a future upgrade of the facility.

In the Fourier domain, the same Kolmogorov turbulence model implies a one--sided OPD power spectral density $S_R(f)\propto f^{-8/3}$ in the inertial range.  When mapped to phase via Eq.~(\ref{eq:phiRF}), this corresponds to $S_\phi(f)\propto f^{-8/3}$ for the atmospheric contribution.  Likewise, an oscillator with fractional frequency noise characterized by an Allan deviation $\sigma_y(\tau)$ produces a phase PSD $S_{\phi,{\rm osc}}(f)$ whose shape is determined by the dominant noise type (e.g.\ white FM, flicker FM); for the hydrogen maser performance assumed in Sec.~\ref{sec:stack}, $S_{\phi,{\rm osc}}(f)$ is well below the atmospheric and photon--noise PSDs over the integration times of interest.  These spectral views are useful when assessing sensitivity to very low--frequency ($\mu$Hz) signatures such as long--period tidal evolution or gravitational waves.

For \(\theta\lesssim0.1^\circ\), the calibrated differential-delay scaling in Eq.~\eqref{eq:diffatm} gives atmospheric contributions in the \(\sim10\)--\(50~\mu\mathrm{m}\) band on \(T\sim10^2~\mathrm{s}\) windows under good-to-excellent conditions. The A/B interleaving cadence is not assumed
to freeze the millisecond-scale optical turbulence; that turbulence is averaged statistically and remains in \(\sigma_{\Delta R,\rm atm}(T,\theta)\). Instead, the cadence requirement is operational: the two reflector measurements must be interleaved rapidly compared with slow differential-delay drift, and the receive schedule must assign each return to the correct transmitted tone and target over the \(\simeq2.56~\mathrm{s}\) round trip. In all operating cases, the differential atmospheric term is therefore carried explicitly in \(C(T)\). In Case~A the total differential range remains photon-floor limited at \(\gtrsim0.1~\mathrm{mm}\), whereas Cases~B and~C can reach the differential bands quoted in Table~\ref{tab:three-cases}.

\subsubsection{Instrument}
\label{sec:instrum-er}

A continuous internal reference through the same RF and ADC chain removes most
common-mode drift. The residual instrument term is specified as a measured
phase-noise or path-length spectrum, not only as a scalar rms number. Let
\(S_{R,\rm inst}(f)\) denote the one-sided PSD of the
internal-reference-corrected path error, normalized so that
\(\mathrm{Var}[x]=\int_0^\infty S_x(f)\,df\) for a stationary scalar process.
For a range average over \(T\),
\begin{equation}
  \sigma^2_{R,\rm inst}(T)
  =
  \int_0^\infty
  S_{R,\rm inst}(f)\,
  |W_0(f;T)|^2\,df,
  \qquad
  W_0(f;T)=
  \frac{1}{T}\int_0^T e^{-i2\pi f t}\,dt .
\label{eq:inst-psd-range}
\end{equation}
For the slope estimator,
\begin{equation}
  \sigma^2_{v,\rm inst}(T)
  =
  \int_0^\infty
  S_{R,\rm inst}(f)\,
  |W_1(f;T)|^2\,df,
  \qquad
  W_1(f;T)=
  \int_0^T
  \frac{12(t-T/2)}{T^3}\,
  e^{-i2\pi f t}\,dt .
\label{eq:inst-psd-slope}
\end{equation}
Equivalently,
\(S_{\phi,\rm inst}(f)=[4\pi f_m/c]^2S_{R,\rm inst}(f)\). At
\(f_m=1~\mathrm{GHz}\), the \(40~\mu\mathrm{m}\) range allocation corresponds to
\[
  \sigma_{\phi,\rm inst}
  =
  \frac{4\pi f_m}{c}(40~\mu\mathrm{m})
  =
  1.68\times10^{-3}~\mathrm{rad}.
\]
Thus the commissioning requirement is
\begin{equation}
  \sigma_{R,\rm inst}(T\simeq100~\mathrm{s})
  \lesssim 40~\mu\mathrm{m},
  \qquad
  \sigma_{\phi,\rm inst}(T\simeq100~\mathrm{s})
  \lesssim 1.7\times10^{-3}~\mathrm{rad}.
\label{eq:instrbudget}
\end{equation}
The station should report \(S_{\phi,\rm inst}(f)\), the overlapping Allan deviation of the internal-reference phase or path, and the integrals in Eqs.~\eqref{eq:inst-psd-range}--\eqref{eq:inst-psd-slope} for the observing windows used in the science reduction. If a two-sided PSD convention is used instead, the corresponding two-sided spectrum must be converted to the
one-sided convention above before applying these equations.

This is the absolute two-way range allocation quoted in the $R_{2\mathrm{w}}$ row of Table~\ref{tab:summary-2w}, and it is realized in practice by the bench, internal reference, and facility parameters summarized in Table~\ref{tab:hw-summary}. The allocations above correspond to the absolute two-way range bands summarized in Sec.~\ref{sec:perf}.

\subsection{Observable-specific noise budgets}
\label{subsec:obs-noise}

For later science analysis it is useful to summarize how the covariance decomposition in Eq.~\eqref{eq:covsplit} maps onto the individual AM--CW observables. Over an integration window $T$ we consider four estimators,
\[
R_{2\mathrm{w}}(T), \qquad v_r(T), \qquad \Delta R_{2\mathrm{w}}(T), \qquad \Delta v_r(T),
\]
corresponding to absolute two--way range, one--way line--of--sight range--rate, and their differential counterparts between nearby CCRs. The corresponding $4\times 4$ covariance matrix $\mathbf{C}(T)$ is defined in Eq.~\eqref{eq:C-decomp-global}; its diagonal entries give the per--observable variances, while the off--diagonal entries encode correlations that can be retained in global fits.

For the design studies in this paper we choose estimators and time origins such that the photon--noise cross covariance between range and range--rate is negligible [cf.\ Eq.~\eqref{eq:covshot}], and we treat the diagonal elements of $\mathbf{C}(T)$ as the primary performance metrics. The representative $1\sigma$ allocations that we adopt for these diagonal elements are collected in Table~\ref{tab:summary-2w}. Each row in that table corresponds to one of the four observables and gives the decomposition of the relevant variance into the photon, atmospheric, instrumental, oscillator, and nonlinearity contributions appearing in Eq.~\eqref{eq:covsplit}. The scalar budgets in Eqs.~\eqref{eq:diffbudget},  \eqref{eq:diffbudget-rr}, and \eqref{eq:budget} are just the range and differential--range rows of Table~\ref{tab:summary-2w}.

\begin{table*}[t]
\centering
\caption{Representative \(1\sigma\) error budget for the four AM--CW observables. Entries are design-level Case~B allocations to the diagonal elements of the covariance matrix \({\bf C}(T)\). Range-rate entries use the corrected slope-estimator scaling in Eq.~\eqref{eq:shot}.}
\label{tab:summary-2w}
\begin{tabular}{l c c l c l}
\hline
Contribution & Symbol &
\(R_{2\mathrm{w}}\) &
\(v_r\) &
\(\Delta R_{2\mathrm{w}}\) &
\(\Delta v_r\) \\
\hline\hline
Photon statistics &
\(\sigma_{\rm shot}\) &
\(30\)--\(55~\mu\mathrm{m}\) &
\(0.2\)--\(0.4~\mu\mathrm{m\,s^{-1}}\) at \(300~\mathrm{s}\) &
\(43\)--\(78~\mu\mathrm{m}\) &
\(1.5\)--\(2.7~\mu\mathrm{m\,s^{-1}}\) at \(100~\mathrm{s}\);\\

 &&&
\(0.2\)--\(0.4~\mu\mathrm{m\,s^{-1}}\) at \(300~\mathrm{s}\) &&
\(0.3\)--\(0.5~\mu\mathrm{m\,s^{-1}}\) at \(300~\mathrm{s}\) \\

Atmosphere (residual)  &
\(\sigma_{\rm atm}\) &
\(50\)--\(150~\mu\mathrm{m}\) &
site and window dependent &
\(10\)--\(30~\mu\mathrm{m}\) &
\(\lesssim\mathrm{few}\times0.1\)--\(1~\mu\mathrm{m\,s^{-1}}\) \\

Instrument (bench, cabling,  &
\(\sigma_{\rm inst}\) &
\(\simeq40~\mu\mathrm{m}\) &
from Eqs.~\eqref{eq:inst-psd-range}--\eqref{eq:inst-psd-slope} &
\(10\)--\(20~\mu\mathrm{m}\) &
from differential internal- \\

metrology) & & & & &
reference PSD \\

Oscillator noise &
\(\sigma_{\rm osc}\) &
\(\lesssim5~\mu\mathrm{m}\) &
\(\ll0.1~\mu\mathrm{m\,s^{-1}}\) &
\(\ll10~\mu\mathrm{m}\) &
\(\ll0.1~\mu\mathrm{m\,s^{-1}}\) \\

Nonlinearity (AM--to--PM, &
\(\sigma_{\rm nl}\) &
\(\lesssim25~\mu\mathrm{m}\) &
\(\ll0.1~\mu\mathrm{m\,s^{-1}}\) &
\(\lesssim10~\mu\mathrm{m}\) &
\(\ll0.1~\mu\mathrm{m\,s^{-1}}\) \\

 multi-tone)  & & & & & \\
\hline
Total (RSS) & $\sigma_{R}$
 &
\(\sim70\)--\(170~\mu\mathrm{m}\); &
\(\lesssim1~\mu\mathrm{m\,s^{-1}}\) for \(T\gtrsim300~\mathrm{s}\) &
\(\sim45\)--\(90~\mu\mathrm{m}\) &
\(\sim0.3\)--\(1~\mu\mathrm{m\,s^{-1}}\) for \\

 &  &
\(\simeq80~\mu\mathrm{m}\) for  &  & &
\(T\simeq300\)--\(1000~\mathrm{s}\) \\

 &  & \((30,60,40)~\mu\mathrm{m}\) &
 & & \\

\hline
\end{tabular}
\end{table*}

The absolute range design point is the favorable Case~B allocation \((\sigma_{R,\rm shot},\sigma_{R,\rm atm},\sigma_{R,\rm inst}) \simeq(30,60,40)~\mu\mathrm{m}\), which gives
\(\sigma_R\simeq78~\mu\mathrm{m}\) in RSS. The broader
\(\sim70\)--\(170~\mu\mathrm{m}\) interval reflects the full Case~B photon and atmosphere ranges.

For the differential observables, atmosphere and instrument are suppressed, but the photon term is larger by \(\sqrt{2}\) for two equal independent links. This lower bound is Eq.~\eqref{eq:diff-photon-floor}. Consequently, the dedicated Case~B differential range band is \(\sim45\)--\(90~\mu\mathrm{m}\)
on \(T\sim100~\mathrm{s}\) windows. The \(20~\mu\mathrm{m}\) level requires the higher flux or longer integration quantified in Eq.~\eqref{eq:diff-20um-flux}. These differential allocations assume separations \(\theta\lesssim0.1^\circ\) and an interleaved A/B sequence whose cadence is short compared with slow differential-delay drift and whose timing model correctly assigns each delayed return to the corresponding transmitted tone and target over the \(\tau_{2\mathrm{w}}\simeq2.56~\mathrm{s}\) lunar round trip. The cadence is not assumed to freeze millisecond-scale optical turbulence; that contribution is averaged statistically and remains in \(\sigma_{\Delta R,\rm atm}(T,\theta)\).

\begin{figure}[t]
  \centering
  \includegraphics[width=0.50 \columnwidth]{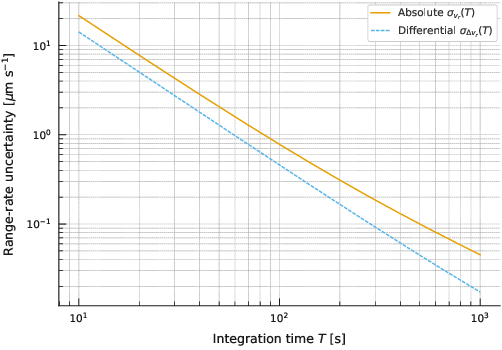}
  \caption{Expected one-way line-of-sight range-rate uncertainties as a function of integration time \(T\) for the absolute observable \(\sigma_{v_r}(T)\) and
the differential observable \(\sigma_{\Delta v_r}(T)\). The photon term uses the centered linear-slope estimator,
\(\sigma_{v_r,\rm shot}=[c/(4\pi f_m)]\sqrt{12}/[T\,{\rm SNR}_{\tt AM}(T)]\). Atmospheric and instrumental drift terms are covariance allocations that must be evaluated from site telemetry and the internal-reference PSD for each
observing sequence.}
  \label{fig:vr-abs-diff}
\end{figure}

For the science analyses it is also helpful to summarize, at a qualitative level, which observables are most directly tied to which classes of physical signals. A compact mapping is shown in Table ~\ref{tab:obs-mapping}.
{}
\begin{table}[h]
\begin{center}
\caption{The science value of different AM-CW LLR observables.}
\label{tab:obs-mapping}
\begin{tabular}{ll}
\hline
Observable & Dominant science drivers \\
\hline\hline
$R_{2\mathrm{w}}(T)$
  & EP tests, $\dot{G}/G$, PPN parameters, long--period GW, secular orbit/libration \\[0.3ex]
$v_r(T)$
  & Near--diurnal tidal signatures, Love numbers, tidal $Q$, short--period core--mantle coupling \\[0.3ex]
$\Delta R_{2\mathrm{w}}(T),\,\Delta v_r(T)$
  & Interior gradients, local tides and loading, regional Love--number variations, \\
  & core--mantle coupling, differential control of station and atmospheric systematics \\
\hline
\end{tabular}
\end{center}
\end{table}

From the standpoint of station systematics, the differential observables are the cleanest: common--mode atmosphere, internal metrology, calibration offsets, and oscillator noise cancel to first order, leaving residuals dominated by photon statistics and the well--understood $\theta^{5/3}$ scaling of the differential atmosphere. Between the two, the differential two--way range $\Delta R_{2\mathrm{w}}(T)$ is especially robust: it depends only on the mean phase difference between nearby CCRs, is insensitive to static path--length offsets, and is limited mainly by well--understood photon noise and the residual differential atmosphere on $\theta \lesssim 0.1^\circ$ separations. Differential range--rate $\Delta v_r(T)$ inherits the same common--mode rejection but is more sensitive to very low--frequency phase noise and any residual modeling error in the modulation--frequency history. In contrast, the absolute observables $R_{2\mathrm{w}}(T)$ and $v_r(T)$ remain indispensable for global parameter estimation, since they carry the full atmospheric and station covariance needed to constrain the long--baseline Earth--Moon dynamics and relativistic parameters.

\subsection{Temporal correlations and multi-epoch covariance}
\label{subsec:temporal-corr}

Throughout this section we have treated ${\bf C}(T)$ as the covariance for a single analysis window, implicitly assuming that different windows are statistically independent once separated by $\gtrsim T$. This approximation is adequate for defining per-window error budgets and implementation gates, but the dominant low-frequency noise sources (Kolmogorov turbulence, bench motion, oscillator flicker) are in reality strongly time--correlated.

Let \(R_{2w}(T_k)\) denote the range estimator formed on a window \(T_k\) with weighting function \(w_k(t)\). If the underlying one-way path fluctuation \(\delta R_{\rm tot}(t)\) has one-sided PSD \(S_R(f)\), using the same normalization as in Eqs.~\eqref{eq:inst-psd-range}--\eqref{eq:inst-psd-slope},
the multi-epoch covariance between two normal points centered at times
\(t_k\) and \(t_{k'}\) can be written, to good approximation, as
\begin{equation}
  {\rm Cov}\!\left[
    R_{2w}(T_k),R_{2w}(T_{k'})
  \right]
  \simeq
  {\rm Re}\!
  \int_0^\infty
  S_R(f)\,
  W_k(f)\,W^{*}_{k'}(f)\,df ,
\label{eq:multi-epoch-cov}
\end{equation}
where \(W_k(f)\) is the Fourier transform of \(w_k(t)\). The spectrum \(S_R(f)\) includes atmospheric, instrumental, oscillator, and any additional colored path-noise contributions. If a two-sided PSD convention is used in an external analysis package, that spectrum must first be converted to the
one-sided convention above before applying Eq.~\eqref{eq:multi-epoch-cov}. For the Kolmogorov atmosphere model of Appendix~\ref{app:atm_turbulence},
\(S_R(f)\propto f^{-8/3}\) in the inertial range, so Eq.~\eqref{eq:multi-epoch-cov} predicts significant correlations between adjacent windows when their centers are separated by \(\lesssim T\).

As a concrete example, using the same Kolmogorov model calibrated to $\sigma_{R,\mathrm{atm}} \simeq 60~\mu\mathrm{m}$ on $T = 100~\mathrm{s}$ windows, a numerical evaluation of the covariance integral shows that two adjacent $100~\mathrm{s}$ windows whose centers are $50~\mathrm{s}$ apart have a correlation coefficient of order $\rho \simeq 0.5$, i.e., such normal points are far from
statistically independent.

In global parameter-estimation or Kalman-filter analyses, these correlations can be included by evaluating Eq.~\eqref{eq:multi-epoch-cov} (or suitable approximations to it) and assembling the full multi-epoch covariance matrix, with the diagonal blocks given by the per-window ${\bf C}(T_k)$ defined in Eq.~\eqref{eq:covsplit}.  At the design stage considered here we use the diagonal elements alone to set hardware and CONOPS requirements, but the formalism above provides a straightforward path to incorporating temporal correlations in science analyses.

\section{Implementation gates}
\label{sec:gates}

The gates in Table~\ref{tab:gates}  provide a concrete mapping between subsystem performance and the overall range error budget in Eq.~(\ref{eq:budget}) and the covariance decomposition in Eq.~(\ref{eq:covsplit}). The SNR requirement on the fine (GHz) tone ensures that the photon--statistical term $\sigma_{R,{\rm shot}}(T)$ remains at or below $0.10$~mm on $T = 30$--$100$~s windows, i.e.\ comfortably below the atmospheric and instrumental allocations. The bound on the rms internal--reference phase directly limits the residual instrumental contribution $\sigma_{R,{\rm inst}}(T)$, while the constraint on $|\dot f_m|$---or explicit correction via Eq.~(\ref{eq:slew})---keeps oscillator--induced range--rate biases far below the target band for $\sigma_{v_r}$. Elevation and seeing cuts control the absolute atmospheric term \(\sigma_{R,\rm atm}(T)\). Differential sequencing controls slow drift and return-path assignment; it is not a frozen-turbulence assumption. The residual differential atmosphere is still evaluated from Eq.~\eqref{eq:diffatm} and carried in \(C(T)\). Together, these gates realize the representative allocations \(\sigma_{R,\rm shot}\), \(\sigma_{R,\rm atm}\), and \(\sigma_{R,\rm inst}\) quoted in Eq.~(\ref{eq:budget}).

\begin{table}[t]
\centering
\caption{Quantitative gates for sub--0.1~mm-class AM--CW LLR.}
\label{tab:gates}
\begin{tabular}{ll}
\hline
Parameter & Performance requirement \\
\hline\hline
Fine-tone SNR &
\({\rm SNR}_{\tt AM}(100~{\rm s})\ge250\) for
\(\sigma_{R,\rm shot}\lesssim0.10~\mathrm{mm}\);
\(\ge800\) for \(\sigma_{R,\rm shot}\lesssim30~\mu\mathrm{m}\) \\
Photon-counting receiver &
time-tag phasor of Eq.~\eqref{eq:photon-phasor};
\(\sigma_t\lesssim50\)--\(100~\mathrm{ps}\) at \(1~\mathrm{GHz}\);
\(\dot N_{\rm det}\tau_{\rm dead}\ll1\) \\
Linear receiver &
RF bandwidth through the precision tone and
\({\rm NEP}_{\rm el}\lesssim(2E_\gamma P_1)^{1/2}\) \\
Doppler derotation &
track \(\Delta f_D=-2f_m v_r/c\);
\(1~\mathrm{km\,s^{-1}}\rightarrow6.67~\mathrm{kHz}\) at \(1~\mathrm{GHz}\) \\
Instrument stability &
\(\sigma_{\phi,\rm inst}(100~{\rm s})\lesssim1.7\times10^{-3}\) rad;
report PSD and overlapping Allan deviation \\
Oscillator drift &
\(|\dot f_m|\lesssim3\times10^{-6}~\mathrm{Hz\,s^{-1}}\) or correct via
Eq.~\eqref{eq:slew} \\
Atmosphere &
\(e\ge30^\circ\); reject poor seeing; carry site-derived
\({\bf C}_{\rm atm}(T)\) in the fit \\

Differential sequencing &
interleaved A/B observing; cadence short compared with slow differential-delay \\

 &
drift, not millisecond optical turbulence; transmit/receive schedule must account  \\

 &
for \(\tau_{2\mathrm{w}}\simeq2.56~\mathrm{s}\) and assign each
return to the correct transmitted tone and target \\

Differential photon floor &
reported \(\sigma_{\Delta R}\) must satisfy
\(\sigma_{\Delta R}\ge
(\sigma^2_{R,A}+\sigma^2_{R,B})^{1/2}\) \\
\hline
\end{tabular}
\end{table}

In practice, these implementation gates can be organized into a simple run--book:
(i) a commissioning phase in which each gate is tested and demonstrated independently (e.g.\ tone SNR on a bright
terrestrial retroreflector, internal--reference stability in closed--loop operation, oscillator characterization);
(ii) an initial operations phase targeting $\sigma_R\simeq 0.2$--0.5\,mm on 10--100\,s windows; and
(iii) a mature phase in which site, hardware, and CONOPS are tuned to approach the 0.05--0.10\,mm design goal and
the differential performance bands in Sec.~\ref{sec:perf}.

\section{Performance summary and recommendations}
\label{sec:perf}

The analysis above supports the following case-dependent performance bands as realistic targets for high-power AM--CW LLR:
\begin{itemize}
\item single-station absolute two-way range: initial \(0.2\)--\(0.5~\mathrm{mm}\); dedicated Case~B design goal \(\sim0.08~\mathrm{mm}\) under favorable atmospheric and instrumental conditions;
\item single-station range-rate: \(\sigma_{v_r}\lesssim1~\mu\mathrm{m\,s^{-1}}\) on several-hundred-second
windows, with \(T\simeq100~\mathrm{s}\) windows generally closer to the \(1\)--\(2~\mu\mathrm{m\,s^{-1}}\) photon floor in Case~B;
\item differential two-way range between nearby CCRs:
\(\sim45\)--\(90~\mu\mathrm{m}\) for dedicated Case~B operation and \(\sim35\)--\(60~\mu\mathrm{m}\) for photon-rich excellent-seeing operation;
\item differential range-rate: \(\sim0.3\)--\(1~\mu\mathrm{m\,s^{-1}}\) on \(T\simeq300\)--\(1000~\mathrm{s}\) windows, depending on flux, atmospheric
residuals, and differential instrument drift.
\end{itemize}

In the notation of Eq.~\eqref{eq:yvec-global}, these four bands are the square roots of the relevant diagonal entries of the \(4\times4\) covariance matrix \(C_{ij}(T)\) in Eq.~\eqref{eq:C-decomp-global}. The allocations in Table~\ref{tab:summary-2w} decompose each diagonal term into photon, atmospheric, instrumental, oscillator, and nonlinearity contributions. The
differential bands are no longer quoted as a universal \(20\)--\(50~\mu\mathrm{m}\) result: the independent photon noise of the two returns imposes the lower bound in Eq.~\eqref{eq:diff-photon-floor}. The \(20~\mu\mathrm{m}\) level is therefore
reserved for the high-flux or longer-integration stretch regime quantified in Eq.~\eqref{eq:diff-20um-flux}.

\paragraph*{Representative operating cases:} It is useful to bundle these assumptions into three representative operating cases that combine photon return, atmospheric residuals, and instrumental stability into a small set of scalar performance targets. Specifically, we take: (i) the three photon--flux regimes of Sec.~\ref{sec:snr}, characterized by detected rates $\dot N_\gamma\simeq (5$--$7)\times10^3$, $(3$--$5)\times10^4$, and $\sim 10^5$~s$^{-1}$; (ii) the three turbulence regimes of Table~\ref{tab:atm-calibration}, with $\sigma_{R,\mathrm{atm}}(T)$ in the $300$--$500$, $50$--$150$, and $30$--$80~\mu$m bands on $T\simeq 30$--$100$~s windows; and (iii) a common instrumental allocation $\sigma_{R,\mathrm{inst}}\simeq 40~\mu$m from Sec.~\ref{sec:instrum-er}. Case~A couples the lowest--flux regime to the generic mid-latitude atmosphere (Regime~A) and therefore reproduces the generic CW concept and error budget of Ref.~\cite{Turyshev:CW-LLR:2025}. Case~B combines the intermediate flux with the dedicated AM--CW turbulence allocation (Regime~B) and defines the baseline design of this work. Case~C couples the photon--rich regime to the excellent--seeing model (Regime~C) and illustrates the performance reachable when both flux and atmosphere are favorable. In Sec.~\ref{sec:hw} we therefore focus the detailed hardware and facility design on Case~B, treating Cases~A and~C primarily as lower and upper performance brackets that are realized by operating the same hardware stack at different photon-flux and turbulence regimes.

\begin{table*}[t]
\caption{Representative operating cases for a high-power AM--CW LLR station
ranging to 10~cm CCRs. Absolute range values are quoted for
\(T\simeq30\)--\(100~\mathrm{s}\). Range-rate values are representative of
\(T\simeq300~\mathrm{s}\) slope fits using Eq.~\eqref{eq:shot}. Differential
ranges include the equal-link photon floor
\(\sqrt{2}\,\sigma_{R,\rm shot}\).}
\label{tab:three-cases}
\begin{tabular}{lc cccc}
\hline
Case &
Detected flux  &
\(\sigma_R\) absolute &
\(\sigma_{v_r}\) absolute &
\(\sigma_{\Delta R}\) differential &
\(\sigma_{\Delta v_r}\) differential \\

 & \(\dot N_\gamma\) [s\(^{-1}\)]& & & \\

\hline\hline
A: generic CW station &
\((5\)--\(7)\times10^{3}\) &
\(0.32\)--\(0.55~\mathrm{mm}\) &
\(\sim0.5\)--\(2~\mu\mathrm{m\,s^{-1}}\) &
\(0.12\)--\(0.20~\mathrm{mm}\) &
\(\sim0.8\)--\(2~\mu\mathrm{m\,s^{-1}}\) \\

B: dedicated AM--CW facility &
\((3\)--\(5)\times10^{4}\) &
\(70\)--\(170\,\mu\mathrm{m}\); \(\simeq80\,\mu\)m  &
\(\sim0.2\)--\(0.6~\mu\mathrm{m\,s^{-1}}\) &
\(45\)--\(90~\mu\mathrm{m}\) &
\(\sim0.3\)--\(0.8~\mu\mathrm{m\,s^{-1}}\) \\

& &  design point & & & \\

C: photon-rich, excellent seeing &
\(\sim10^{5}\) &
\(55\)--\(95~\mu\mathrm{m}\) &
\(\sim0.1\)--\(0.3~\mu\mathrm{m\,s^{-1}}\) &
\(35\)--\(60~\mu\mathrm{m}\) &
\(\sim0.2\)--\(0.5~\mu\mathrm{m\,s^{-1}}\) \\
Stretch: high-flux differential &
\(\gtrsim2.3\times10^{5}\) &
case dependent &
case dependent &
\(\sim20\)--\(40~\mu\mathrm{m}\) &
case dependent \\
\hline
\end{tabular}
\end{table*}

Table~\ref{tab:three-cases} separates the generic CW link from the dedicated and photon-rich AM--CW regimes. Case~A reproduces the earlier 1~kW CW concept on a conventional 1~m platform, but its differential range is not in the tens-of-micrometers regime because the two-link photon floor already exceeds \(0.1~\mathrm{mm}\). Case~B corresponds to a purpose-built AM--CW facility at a good site, with higher throughput, explicit elevation and SNR cuts, continuous internal referencing, calibrated RF/optical paths, and a detector
back end satisfying either Eq.~\eqref{eq:photon-phasor} or
Eq.~\eqref{eq:linear-nep}. Under the favorable allocation
\((30,60,40)~\mu\mathrm{m}\), the absolute range RSS is
\(\simeq80~\mu\mathrm{m}\); over the full atmospheric and photon intervals, the Case~B range is broader, \(\sim70\)--\(170~\mu\mathrm{m}\). Case~C mainly provides photon margin once atmosphere and instrument terms are controlled. Differential ranges near \(20~\mu\mathrm{m}\) require the stretch flux or longer integration given in Eq.~\eqref{eq:diff-20um-flux}.

At \(f_m=1~\mathrm{GHz}\), \(c/(4\pi f_m)=2.38567\times10^{-2}~\mathrm{m\,rad^{-1}}\). Reaching \(\sigma_R=0.10~\mathrm{mm}\) requires \(\sigma_\phi\simeq4.2\times10^{-3}~\mathrm{rad}\), or
\({\rm SNR}_{\tt AM}\simeq240\). The nominal Case~B photon rates give \({\rm SNR}_{\tt AM}(100~\mathrm{s})\simeq(4\)--\(8)\times10^2\), corresponding to \(\sigma_{R,\rm shot}\simeq30\)--\(55~\mu\mathrm{m}\). Values \({\rm SNR}_{\tt AM}\gtrsim10^3\) on \(100~\mathrm{s}\) windows correspond to
Case~C or the stretch high-flux regime, not to every 1~kW/1~m implementation. Thus photon statistics are not the only limiting term in Cases~B/C, but they remain a significant floor in Case~A and in differential two-link observables.

Using (\ref{eq:budget}), a representative decomposition of the absolute two-way range variance ($R_{2\mathrm{w}}$ column of Table~\ref{tab:summary-2w}) is
\begin{equation}
\label{eq:errr-sum}
\sigma_{R,{\rm shot}}\sim 30\,\mu{\rm m},\quad
\sigma_{R,{\rm atm}}\sim 60\,\mu{\rm m},\quad
\sigma_{R,{\rm inst}}\sim 40\,\mu{\rm m},
\end{equation}
which yields
\begin{equation}
\sigma_R \sim \sqrt{30^2+60^2+40^2}\ \mu{\rm m} \approx 80\ \mu{\rm m}.
\label{eq:err-sum-rss}
\end{equation}
This illustrates that sub--0.1\,mm two--way precision is realistic as a design goal, provided atmospheric variation and instrument residuals are systematically held below the $\sim 50$--$100\,\mu$m level.

The oscillator and nonlinearity terms ${\bf C}_{\rm osc}(T)$ and ${\bf C}_{\rm nl}(T)$ from Eq.~(\ref{eq:covsplit}) are implicitly included in this allocation. As discussed in Sec.~\ref{sec:stack}, the assumed maser performance implies $\sigma_{R,{\rm osc}} \lesssim 5~\mu\mathrm{m}$ [Eq.~(\ref{eq:sigma_R_osc})], and the AM--to--PM and multi-tone nonlinearity constraints keep $\sigma_{R,{\rm nl}} \lesssim 25~\mu\mathrm{m}$ [Eq.~(\ref{eq:sig_r})].  Both are therefore comfortably absorbed within the $\sim 40~\mu\mathrm{m}$ instrumental term in Eq.~(\ref{eq:budget}), confirming that the dominant residuals are atmospheric and bench-related.

For range-rate, Eq.~\eqref{eq:shot} gives the photon-limited slope precision. At \(f_m=1~\mathrm{GHz}\) and \(T=100~\mathrm{s}\), the Case~B photon rates imply one-second phase uncertainties
\[
  \sigma_{\phi,1{\rm s}}
  \simeq
  \frac{1}{(a_m/2)\sqrt{\dot N_\gamma}}
  \simeq
  0.013\mbox{--}0.023~\mathrm{rad}.
\]
The corresponding photon-limited range-rate floor is
\begin{equation}
  \sigma_{v_r,\rm shot}(100~\mathrm{s})
  =
  \frac{c}{4\pi f_m}
  \frac{\sqrt{12}}{100~\mathrm{s}\,{\rm SNR}_{\tt AM}(100~\mathrm{s})}
  \simeq
  1.1\mbox{--}1.9~\mu\mathrm{m\,s^{-1}}
\label{eq:rr-numeric}
\end{equation}
for Case~B. At \(T=300~\mathrm{s}\) this becomes
\(0.20\)--\(0.37~\mu\mathrm{m\,s^{-1}}\), and at \(T=1000~\mathrm{s}\) it becomes \(0.03\)--\(0.06~\mu\mathrm{m\,s^{-1}}\), before adding atmospheric
and instrumental drift covariance. Thus sub-\(\mu\mathrm{m\,s^{-1}}\) range-rate is a several-hundred-second Case~B claim, or a shorter-window claim only in photon-rich operation.

In differential mode, common-mode station and atmospheric terms are suppressed, but the independent photon noise from the two reflector returns remains. For equal links the photon contribution is \(\sigma_{\Delta R,\rm shot}=\sqrt{2}\,\sigma_{R,\rm shot}\), giving \(43\)--\(78~\mu\mathrm{m}\) for Case~B at \(T=100~\mathrm{s}\) before adding residual differential atmosphere or instrument terms. Adding representative
\(\sigma_{\Delta R,\rm atm}\simeq10\)--\(30~\mu\mathrm{m}\) and \(\sigma_{\Delta R,\rm inst}\simeq10\)--\(20~\mu\mathrm{m}\) gives \(\sigma_{\Delta R}\simeq45\)--\(90~\mu\mathrm{m}\) for the dedicated Case~B facility and \(\sim35\)--\(60~\mu\mathrm{m}\) for Case~C. Applying the corrected slope-fit logic to \(\Delta\phi(t)\) yields sub-\(\mu\mathrm{m\,s^{-1}}\)
differential range-rate most naturally on \(T\simeq300\)--\(1000~\mathrm{s}\) windows. The \(20\)--\(40~\mu\mathrm{m}\) band should be reported only as a stretch case requiring the flux or integration time in Eq.~\eqref{eq:diff-20um-flux}.

\section{Hardware and CONOPS}
\label{sec:hw}

\begin{figure}[t]
\centering
\begin{tikzpicture}[
  >=angle 90,           
  node distance=0.9cm,
  block/.style={
    draw,
    rectangle,
    rounded corners,
    align=center,
    minimum height=0.9cm,
    text width=7.6cm
  },
  dashedblock/.style={
    draw,
    rectangle,
    rounded corners,
    dashed,
    align=center,
    minimum height=1.2cm,
    text width=3.5cm
  },
  line/.style={->,thick}
]

\node[block] (maser) {Hydrogen maser frequency reference\\
RF synthesis of tones $\{f_{m,i}\}$ (50\,MHz--1\,GHz)\\
short-term stability $\sigma_y(\tau)$};

\node[block,below=of maser] (modamp) {Modulation \& power amplification\\
LiNbO$_3$ MZM, AM index $a_m$ at $\{f_{m,i}\}$\\
single-frequency 1064\,nm seed, MOPA to $P_0\sim 1$~kW};

\node[block,below=of modamp] (telmoon) {Tx/Rx telescope and lunar path\\
aperture $D\simeq 1$~m, pointing and tracking\\
10~cm CCRs, round-trip time $\tau\approx 2.56$~s, link budget $\dot N_\gamma$};

\node[block,below=of telmoon] (rxchain) {Receive optics, detector, RF front-end, ADC\\
spectral/spatial filtering, SNSPD / InGaAs\\
RF envelope sampling at the modulation tones};

\node[block,below=of rxchain] (dspest) {Digital lock--in and estimators\\
per-tone phases $\phi_i(t)$, lock--in SNR ${\rm SNR}_{\tt AM}(T)$\\
$R_{2\mathrm{w}}(T),\,v_r(T),\,\Delta R_{2\mathrm{w}}(T),\,\Delta v_r(T)$, covariance ${\bf C}(T)$};

\draw[line] (maser) -- (modamp);
\draw[line] (modamp) -- (telmoon);
\draw[line] (telmoon) -- (rxchain);
\draw[line] (rxchain) -- (dspest);

\node[
  dashedblock,
  right=2.0cm of rxchain
] (iref) {Internal reference path\\
short, co-routed optical/RF loop\\
measured through same RF \& ADC chain\\
internal phase $\phi_{\rm inst}(t)$};

\path (iref.north west) -- (iref.south west)
  coordinate[pos=0.33] (iref_in)
  coordinate[pos=0.67] (iref_out);

\coordinate (knee_x) at ($(rxchain.east)!0.5!(iref.west)$);


\coordinate (entry_knee1) at ($(knee_x |- modamp.east)$); 
\coordinate (entry_knee2) at ($(knee_x |- iref_in)$);      
\draw[line,dashed] (modamp.east) -- (entry_knee1) -- (entry_knee2) -- (iref_in);

\coordinate (iref_out_left) at ($(knee_x |- iref_out)$);    
\coordinate (exit_down)     at ($(knee_x |- dspest.east)$); 

\draw[thick,dashed] (iref_out) -- (iref_out_left);  
\draw[thick,dashed] (iref_out_left) -- (exit_down); 
\draw[line,dashed] (exit_down) -- (dspest.east);    

\end{tikzpicture}
\caption{
Simplified schematic of the RF-coherent high-power AM--CW LLR station. A hydrogen maser and RF synthesis stage generate the modulation tones $\{f_{m,i}\}$, which drive a LiNbO$_3$ MZM and MOPA chain to impose amplitude modulation with index $a_m$ on a 1064\,nm CW beam at power $P_0\sim 1$~kW. The transmit/receive telescope launches the beam to the lunar 10~cm CCRs and collects the weak, delayed return, which is filtered, detected, and digitized by the receive chain. Digital lock--in processing and estimators on the precision tone, aided by a short internal reference path that supplies $\phi_{\rm inst}(t)$, produce the observables $R_{2\mathrm{w}}(T)$, $v_r(T)$, $\Delta R_{2\mathrm{w}}(T)$, and $\Delta v_r(T)$.
}
\label{fig:amcw_block}
\end{figure}
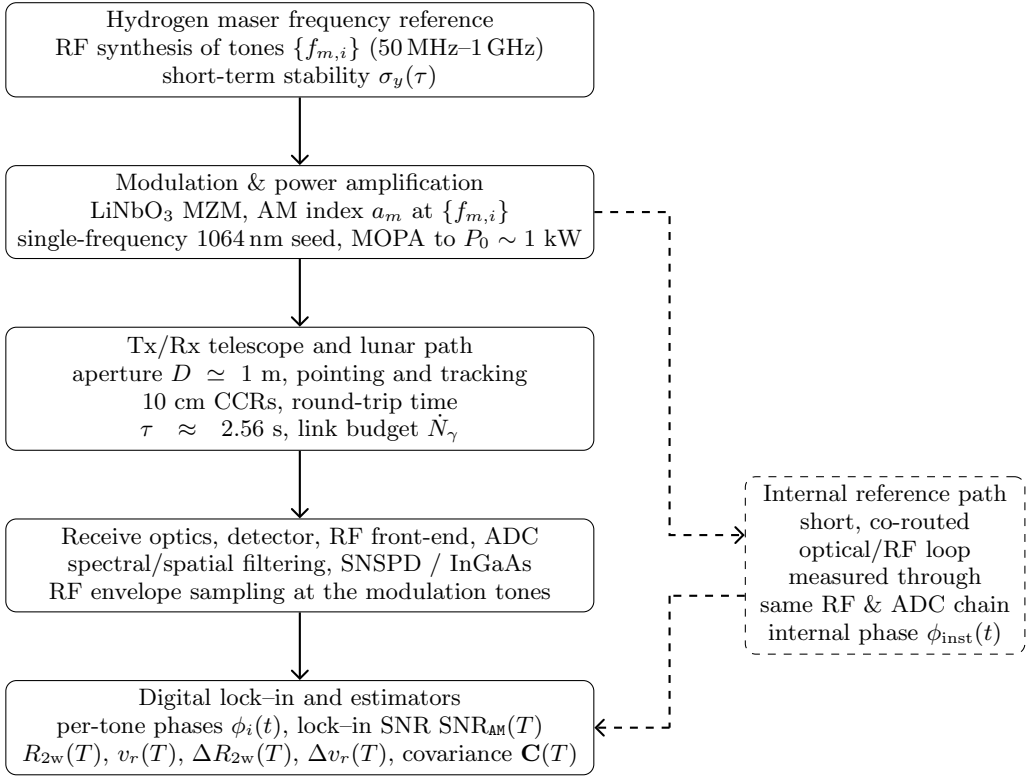

The AM--CW LLR station behaves as an RF interferometer coupled to a high-power optical transmitter/receiver, as summarized in Fig.~\ref{fig:amcw_block}. The hydrogen maser defines the frequency reference for the modulation tones $\{f_{m,i}\}$, the LiNbO$_3$ modulator and MOPA chain realize a kW-class 1064\,nm CW beam with amplitude index $a_m$, and the lunar CCRs return a delayed, attenuated replica of the imposed modulation. The internal reference path, measured through the same RF and ADC chain, provides the phase $\phi_{\rm inst}(t)$ used to remove common-mode instrument drift before forming the range and range--rate estimators defined in Sec.~\ref{sec:model}.

In the notation of Eq.~\eqref{eq:covsplit}, the blocks in Fig.~\ref{fig:amcw_block} map most directly onto the oscillator, photon, and instrumental contributions ${\bf C}_{\rm osc}(T)$, ${\bf C}_{\rm shot}(T)$, and ${\bf C}_{\rm inst}(T)$ that set the precision of the AM--CW observables. In this section we describe the subsystems and the operational concept at a level that supports concrete requirements.

In the remainder of this section we specialize to the dedicated AM--CW facility (Case~B in Table~\ref{tab:three-cases}), characterized by the baseline link parameters in Table~\ref{tab:link-budget} and the hardware summary in
Table~\ref{tab:hw-summary}. In Cases A and C the same hardware stack is operated at different photon-flux and turbulence regimes (Table~\ref{tab:three-cases}), so a separate hardware table is not required; the performance deltas are driven by site quality, throughput, and observing strategy rather than qualitatively different subsystems.

\begin{table}[t]
  \centering
  \caption{Baseline AM--CW LLR station hardware and facility parameters for the
  dedicated facility (Case~B).  Values are design targets consistent with the link
  budget in Table~\ref{tab:link-budget}, the error allocations in
  Table~\ref{tab:summary-2w}, and the covariance decomposition
  $\Cov(T)=\Cshot+\Catm+\Cinst+\Cosc+\Cnl$ in Eq.~\eqref{eq:covsplit}.}
  \label{tab:hw-summary}
  \begin{tabular}{lll}
    \hline
    Subsystem & Quantity & Design value / requirement \\
    \hline\hline
    Frequency reference &
    Oscillator type &
    Hydrogen maser or equivalent ultra--stable reference \\
    &
    Allan deviation $\sigma_y(\tau)$ &
    $\lesssim 10^{-14}$ for $\tau=1$--$10~\mathrm{s}$,
    $\lesssim 10^{-15}$ at $\tau\sim 10^{2}~\mathrm{s}$ \\
    &
    Range contribution $\sigma_{R,{\rm osc}}$ &
    $\lesssim 5~\mu\mathrm{m}$ on $T\sim 10$--$10^{2}~\mathrm{s}$ 
    [Eqs.~\eqref{eq:sigma_R_osc}--\eqref{eq:sigma_R}] \\[0.4em]
    Laser and modulation &
    Optical wavelength $\lambda$ &
    $1064~\mathrm{nm}$ (single-frequency seed) \\
    &
    Mean optical power $P_0$ &
    $1~\mathrm{kW}$ CW at the telescope aperture \\
    &
    AM index $a_m$ on precision tone &
    $0.3$--$0.7$ \\
    &
    Tone set $\{f_{m,i}\}$ &
    $\{50,\,50.1,\,200,\,1000\}~\mathrm{MHz}$ (illustrative) \\
    &
    Beam quality $M^{2}$ &
    $\lesssim 1.3$ at full power \\
    &
    AM$\rightarrow$PM residual $|\delta\phi_{\rm AM\rightarrow PM}|$ &
    $\ll 10^{-3}~\mathrm{rad}$ on $T\simeq 10$--$100~\mathrm{s}$ \\
    &
    Nonlinearity range term $\sigma_{R,{\rm nl}}$ &
    $\lesssim 25~\mu\mathrm{m}$ at $f_m=1~\mathrm{GHz}$
    [Eqs.~\eqref{eq:sigmaR-nl}--\eqref{eq:phi-AMPM-req}] \\[0.4em]
    Tx/Rx optics &
    Telescope aperture $D$ &
    $1$--$2~\mathrm{m}$ (monostatic or near-monostatic) \\
    &
    Optical bandpass &
    $\sim 1$--$3~\mathrm{nm}$ at $1064~\mathrm{nm}$ \\
    &
    Lunar CCRs &
    $d_{\rm CCR}=10~\mathrm{cm}$ next-generation arrays \\[0.4em]
       Hardware regime 
         & 1 m, $\eta_{\rm eff} \approx 0.2$, & Generic seeing \\
         & 1--2 m, $\eta_{\rm eff} \approx 0.35$--$0.4$, & Good site \\
         & 1--2 m, $\eta_{\rm eff} \gtrsim 0.4$, & Excellent site \\[0.4em]

     Detector and back-end &
    Detector option~1 &
    SNSPD or equivalent time-tagging detector at  \\
    
 & & \(1064~\mathrm{nm}\);
    system efficiency \(\gtrsim0.2\), dark counts  \\
  & & 
    \(\lesssim10~\mathrm{s^{-1}}\), single-event timing jitter
    \(\lesssim50\)--\(100~\mathrm{ps}\),  \\   
  & &  and
    \(\dot N_{\rm det}\tau_{\rm dead}\ll1\) \\   
    
    &
    Detector option~2 &
    Low-noise InGaAs or optically assisted linear  \\
    
     & &
    receiver at higher flux;
    analog RF bandwidth  \\ 
    
         & &
    through the precision tone and
    \({\rm NEP}_{\rm el}\lesssim(2E_\gamma P_1)^{1/2}\) \\   
    
    &
    Phase extraction &
    Photon-counting mode uses the derotated time-tag  \\
 
    && phasor
    \(Z_m=\sum_n\exp[-i\Phi_{\rm LO}(t_n)]\);
    linear mode uses  \\ 
    
    &&  analog I/Q demodulation \\ 
    
    &
    ADC / time-tagger clock stability &
    untracked differential timing jitter
    \(\lesssim0.27~\mathrm{ps}\) rms for \\
    
     & &
  a \(40~\mu\mathrm{m}\)
    allocation at \(1~\mathrm{GHz}\);
    \(0.5~\mathrm{ps}\) corresponds \\
    
      & &
 to \(75~\mu\mathrm{m}\) and is acceptable only if common-mode,  \\   
 
       & &
calibrated, or averaged without phase bias \\   
       
    &
    Dynamic range / resolution &
    Sufficient that amplitude fluctuations do not bias \\
    
        & &
phase estimates at the    \(10^{-3}\)~rad level \\
       
    Internal reference &
    Reference geometry &
    Short, co-routed optical/RF loop through \\
    && same RF+ADC chain \\
    &
    Residual phase rms$\{\phi_{\rm inst}\}$ &
    $\lesssim 2\times 10^{-3}~\mathrm{rad}$ over $T\simeq 100~\mathrm{s}$
    ($\Rightarrow \sigma_{R,{\rm inst}}\lesssim 40~\mu\mathrm{m}$) \\[0.4em]
    Facility / environment &
    Laser wall-plug efficiency &
    $\sim 20$--$30\%$ (electrical draw $\sim 3$--$5~\mathrm{kW}$ for laser) \\
    &
    Temperature stability $\delta T$ &
    $\lesssim 0.1~\mathrm{K}$ over $T\simeq 100~\mathrm{s}$ in laser and RF rooms \\
    &
    Bench thermal path term $\sigma_{R,{\rm inst,th}}$ &
    $\lesssim 10~\mu\mathrm{m}$ on $T\simeq 100~\mathrm{s}$ [Eq.~\eqref{eq:thermal-allocation}] \\
    &
    Vibrational path term $\sigma_{R,{\rm inst,mech}}$ &
    $\lesssim 20~\mu\mathrm{m}$ on $T\sim 10$--$100~\mathrm{s}$ [Eq.~\eqref{eq:mech-allocation}] \\
    &
    Total instrumental allocation $\sigma_{R,{\rm inst}}$ &
    $\simeq 40~\mu\mathrm{m}$ (design value in Table~\ref{tab:summary-2w}) \\
    \hline
  \end{tabular}
\end{table}

\subsection{Common hardware stack}
\label{sec:stack}

\subsubsection{Frequency reference} 

A hydrogen maser (or equivalent ultra--stable oscillator)~\cite{Allan1966ProcIEEE}.  It synthesizes the modulation tones $\{f_m\}$ in the 50\,MHz--1\,GHz range and provides timing for the ADCs. Short--term fractional frequency instability (Allan deviation) at the level of $10^{-14}$ on $1$--$10$~s and better than $10^{-15}$ on $10^2$~s ensures that stochastic oscillator noise is negligible in the range and range--rate error budgets. A simple bound on the two-way range contribution over the round-trip light time $\tau_{2\mathrm{w}} \simeq 2.56$~s is
\begin{equation}
\sigma_{R,{\rm osc}} \simeq \frac{c\,\tau_{2\mathrm{w}}}{2}\,\sigma_y(\tau_{2\mathrm{w}}),
\label{eq:sigma_R_osc}
\end{equation}
where $\sigma_y(\tau)$ is the Allan deviation of the maser. For $\sigma_y(1\text{--}10~\mathrm{s}) \lesssim 10^{-14}$ one finds $\sigma_{R,{\rm osc}} \lesssim 4~\mu\mathrm{m}$, so that $C_{\rm osc}(T)$ in Eq.~(\ref{eq:covsplit}) is safely subdominant to the atmospheric and instrumental contributions. The corresponding range--rate contribution on a window $T$ scales as
\begin{equation}
\sigma_{v_r,{\rm osc}}(T) \sim \frac{\sigma_{R,{\rm osc}}}{T} \ll 1~ \mu\mathrm{m\,s^{-1}}
\end{equation}
for $T \gtrsim 10$~s, consistent with the target band for $\sigma_{v_r}$ in Sec.~\ref{sec:perf}.

In the notation of Eq.~(\ref{eq:covsplit}), these bounds imply a scalar oscillator contribution
\begin{equation}
  \sigma_{R,\mathrm{osc}}(T) \equiv \sqrt{C_{\mathrm{osc},11}(T)} \lesssim 5~\mu\mathrm{m}
  \label{eq:sigma_R}
\end{equation}
on $T \sim 10$--$10^2$~s windows. This is more than an order of magnitude below the atmospheric and instrumental allocations adopted in Sec.~\ref{sec:errors}, and it justifies treating ${\bf C}_{\mathrm{osc}}(T)$ as a sub-dominant term in the total covariance decomposition of Eq.~(\ref{eq:covsplit}).

\subsubsection{Laser and modulation}

A single--frequency 1064\,nm master oscillator feeding a master--oscillator power--amplifier (MOPA) chain. The modulation is imposed at the seed using a LiNbO\(_3\) Mach--Zehnder modulator with tone set \(\{f_{m,i}\}\); typical amplitude--modulation indices \(a_m \sim 0.3\text{--}0.7\) are feasible. The MOPA is operated well below saturation to preserve AM index and linear phase transfer. High--power optical isolators protect the seed. The amplifier chain is designed for near--diffraction--limited beam quality ($M^2\lesssim 1.3$), minimal pointing jitter, and stable polarization into the telescope aperture~\cite{Degnan1993ProcIEEE,Turyshev:CW-LLR:2025}.

Characterization of the LiNbO$_3$ modulator response at the operating tones is essential to control AM--to--PM
conversion and harmonic distortion.  Residual AM--to--PM conversion on the precision tone\footnote{In practice we designate one of the modulation tones, typically at $f_m = 1$\,GHz, as the ``precision tone'' whose phase is used to form the primary range and range--rate estimators,
while lower--frequency tones serve primarily for ambiguity resolution and diagnostics.} can be parameterized as an equivalent phase perturbation $\delta\phi_{\rm AM\rightarrow PM}(t)$ entering $\delta\phi_{\rm nl}(t)$ in Eq.~(\ref{eq:meas}). For a given modulation index, the corresponding nonlinearity-induced range error on a window $T$ is
\begin{equation}
\label{eq:sigmaR-nl}
\sigma_{R,{\rm nl}}(T) \simeq \frac{c}{4\pi f_m}\,\mathrm{rms}\{\delta\phi_{\rm AM\rightarrow PM}\},
\end{equation}
so that enforcing
\begin{equation}
\label{eq:phi-AMPM-req}
\mathrm{rms}\{\delta\phi_{\rm AM\rightarrow PM}\} \lesssim 10^{-3}\,\mathrm{rad}
\end{equation}
at $f_m = 1~\mathrm{GHz}$ limits $\sigma_{R,{\rm nl}}$ to $\lesssim 25~\mu\mathrm{m}$. This contribution is collected in ${\bf C}_{\rm nl}(T)$ in Eq.~(\ref{eq:covsplit}) and is therefore comfortably below the atmospheric and instrumental allocations in Eq.~(\ref{eq:budget}).

In practice we require residual phase modulation at $f_m$ induced by nominal AM to satisfy
\begin{equation}
  |\delta\phi_{\rm AM\rightarrow PM}|\ll 10^{-3}\ {\rm rad}
\end{equation}
over $T\simeq 10$--100\,s, corresponding to a two--way range bias $\ll 25\,\mu{\rm m}$ at $f_m=1$\,GHz.  Harmonic content of the AM envelope at $2f_m$ and higher is kept below ${\sim}-40$\,dBc, so that the digital lock--in---which assumes a single dominant tone---does not acquire systematic phase shifts from higher harmonics.  Residual effects are included in the nonlinearity term $\delta\phi_{\rm nl}(t)$ in Eq.~(\ref{eq:meas}) and contribute to $\mathbf{C}_{\rm nl}$ in Eq.~(\ref{eq:covsplit}).

\subsubsection{Transmit/receive optics}

A 1--2\,m telescope, used in a monostatic or near--monostatic configuration. A fast steering mirror provides milli--arcsecond pointing control. A narrow optical passband (on the order of 1--3\,nm) and spatial filtering in the focal plane suppress lunar background, especially during bright phases. Aperture shape and central obscuration are chosen to minimize far--field sidelobes and to maintain good encircled energy at the reflector arrays; coating design minimizes ghost reflections that could contaminate the lock--in.

The assumed 10~cm CCRs are consistent with proposed next-generation lunar retroreflectors~\cite{Currie2011Acta,Turyshev:CW-LLR:2025,Turyshev-CCR:2025}, which motivate the use of kW-class CW transmitters and GHz-class modulation frequencies.

\subsubsection{Detector and back-end}

Two receiver implementations are possible. In the photon-counting implementation, an SNSPD or equivalent detector at 1064~nm time-tags individual photon arrivals relative to the maser-referenced modulation phase. The detector is not used as a GHz analog photodiode; the RF phase is recovered from the arrival-time phasor in Eq.~\eqref{eq:photon-phasor}. The relevant requirements are detection efficiency, dark count, single-event timing jitter, dead time, count-rate margin, and time-tag clock stability. The detected rates in Cases A--C are \(\lesssim10^5~\mathrm{s^{-1}}\), well below a
\(10~\mathrm{MHz}\) saturation scale, so dead time is not a fundamental limitation provided \(\dot N_{\rm det}\tau_{\rm dead}\ll1\). Timing jitter enters through the contrast factor in Eq.~\eqref{eq:jitter-contrast}.

In the linear implementation, the receiver is a low-noise InGaAs photodiode or an optically assisted receiver followed by RF I/Q demodulation. This option requires analog bandwidth through the highest modulation tone and an RF-band equivalent input noise satisfying Eq.~\eqref{eq:linear-nep}. For the fW powers
of Case~B, shot-limited linear operation requires
\({\rm NEP}_{\rm el}\) of order \(5\times10^{-17}~\mathrm{W\,Hz^{-1/2}}\). Both receiver modes produce the same data products: per-tone complex phasors, phase estimates, phase slopes, lock-in SNRs, and covariance estimates.

\subsubsection{Internal reference}

A short, stable reference path is measured continuously through the same RF and ADC chain. Its phase $\phi_{\rm inst}(t)$ monitors instrument drift: laser/mixer phase noise, RF path length, ADC timing, and residual AM--to--PM conversion. Subtracting $\phi_{\rm inst}(t)$ from the lunar phase per tone removes the bulk of instrument path and electronics drift. Residual instrument phase noise on the precision tone is required to remain below a few$\times 10^{-3}$\,rad over 100\,s; bench design and cable routing are optimized to satisfy this constraint.

\subsection{Facility-level infrastructure for high-power operation}
\label{sec:facility}

Realizing a 1\,kW--class AM--CW station requires facility infrastructure that preserves the path--length stability
implicit in the precision goals while handling kilowatt--level optical and several--kilowatt electrical loads.

\subsubsection{Thermal and electrical budget}  

A 1\,kW optical transmitter with overall wall--plug efficiency of order 20--30\% implies an electrical draw of ${\sim}3$--5\,kW for the laser system alone, plus additional load for chillers, RF electronics, and cryogenics.  The laser room and RF bench are therefore designed with several kilowatts of heat rejection and active temperature control that keeps bulk temperature excursions within ${\sim}0.1$\,K over 100\,s. Using an effective optical path length $L \simeq 1~\mathrm{m}$ and an effective coefficient of thermal expansion $\alpha_{\rm eff}$ for the metrology bench, the residual instrument path fluctuation
\begin{equation}
  \delta R_{\rm inst} \simeq \alpha_{\rm eff} L\,\delta T
\end{equation}
must satisfy $\delta R_{\rm inst} \ll 10~\mu\mathrm{m}$ over $T\simeq 100~\mathrm{s}$ in order to
respect the phase constraint stated above, implying $\alpha_{\rm eff}\,\delta T \lesssim 10^{-5}$.
In terms of the scalar error budget in Eq.~\eqref{eq:budget} and the covariance decomposition
in Eq.~\eqref{eq:covsplit}, this condition corresponds to an allocation
\begin{equation}
  \sigma_{R,\mathrm{inst,th}}(T \simeq 100~\mathrm{s})
  \;\lesssim\; 10~\mu\mathrm{m},
  \label{eq:thermal-allocation}
\end{equation}
for the bench and thermal contributions to $R_{\rm inst}(t)$, i.e.\ a contribution $C_{\mathrm{inst},11}(T\simeq 100~\mathrm{s}) \lesssim (10~\mu\mathrm{m})^2$, leaving additional margin within $\mathbf{C}_{\rm inst}(T)$ for residual electronic, alignment, and calibration terms.

\subsubsection{Mechanical and vibrational environment}  

The RF/optical bench and beam transport to the telescope are mounted on a low--vibration pier, with the amplifier chain, modulator, and internal reference arranged to minimize differential path motion between the lunar and reference  channels. Vibration from cryocoolers and HVAC systems is isolated to keep induced path--length noise below the atmospheric floor on the relevant time scales. Expressed in the notation of Eq.~\eqref{eq:budget}, the goal is to keep the vibrational contribution to the instrumental variance at the level
\begin{equation}
  \sigma_{R,\mathrm{inst,mech}}(T) \;\lesssim\; 20~\mu\mathrm{m},
  \qquad T \sim 10\text{--}100~\mathrm{s},
  \label{eq:mech-allocation}
\end{equation}
so that, combined with the thermal allocation above, the total $\sigma_{R,\mathrm{inst,res}}(T)$ remains within the $\sim 40~\mu\mathrm{m}$ budget adopted in Sec.~\ref{sec:perf}.

The $\sim 10~\mu\mathrm{m}$ target in Eq.~\eqref{eq:thermal-allocation} applies specifically to bench and thermal contributions to $R_{\rm inst}(t)$.
Together with the $\sim 20~\mu\mathrm{m}$ vibrational allocation in Eq.~\eqref{eq:mech-allocation}, and allowing $\sim 10$--$20~\mu\mathrm{m}$ of headroom for residual electronic, alignment, and calibration terms, this is compatible with the overall $\sigma_{R,\mathrm{inst}}\simeq 40~\mu\mathrm{m}$ budget in Table~\ref{tab:summary-2w}.

\subsubsection{Telescope enclosure and site monitoring}  

The telescope sits in a dome or roll--off enclosure designed to minimize local seeing and thermal gradients across the primary.  Environmental sensors (temperature, pressure, humidity, wind) and a seeing monitor (e.g.\ MASS/DIMM) provide per--night estimates of Fried parameter $r_0$, coherence time $\tau_0$, and effective differential-delay angular scale $\theta_0$.  These feed into the atmospheric term $\sigma_{R,{\rm atm}}(T)$ and into the quality cuts applied in the observing sequence.

Quantitatively, for the Kolmogorov structure function adopted later for the atmospheric contribution to the covariance, the combination of site selection, elevation cuts, and turbulence monitoring is designed to enforce
$\sigma_{R,\mathrm{atm}}(T) \lesssim (5$--$10)\times 10^{-5}\,\mathrm{m}$
on $T = 30$--$100~\mathrm{s}$, consistent with the allocations in the precision budget.

\subsubsection{High-power safety and beam control}

Operation of a 1\,kW, 1064\,nm CW beam requires a safety system with hardwired shutters, interlocked doors, and software limits on telescope pointing.  The control system enforces night--time operation only, checks for aircraft and satellite exclusion zones, and closes shutters in the event of any interlock violation.  Nominal ocular--hazard distances are tens of kilometers, so these engineered controls are essential for safe routine operation and to limit illumination of the lunar surface outside the intended reflector fields.

\subsubsection{Timing and frequency distribution}

The hydrogen maser reference is distributed to the modulation source, local oscillators, time taggers, and ADC clocks over phase-stable RF or optical links. An untracked timing error \(\delta t\) at the digitizer or time tagger
produces \(\delta\phi=2\pi f_m\delta t\) and therefore
\begin{equation}
  \delta R
  =
  \frac{c}{4\pi f_m}\delta\phi
  =
  \frac{c\,\delta t}{2}.
\end{equation}
Thus \(0.5~\mathrm{ps}\) corresponds to
\(\delta R\simeq75~\mu\mathrm{m}\), not \(\ll0.1~\mathrm{mm}\). To keep untracked timing jitter below a \(40~\mu\mathrm{m}\) instrument allocation requires \(\delta t\lesssim0.27~\mathrm{ps}\) rms; to keep it below \(25~\mu\mathrm{m}\) requires \(\delta t\lesssim0.17~\mathrm{ps}\) rms. Larger raw jitter is acceptable only when it is common-mode between the lunar
and internal-reference channels, explicitly calibrated, or statistically averaged without phase bias. In photon-counting mode this differential clock stability is separate from the single-event detector jitter, which enters through Eq.~\eqref{eq:jitter-contrast}.

\subsubsection{Geophysical co-location and ancillary sensors}
\label{sec:geophysics}

For long-term interpretation of sub-mm normal points it is advantageous to embed the AM--CW LLR station in a broader geodetic and gravimetric environment. Co-located continuous Global Navigation Satellite System (GNSS), a superconducting gravimeter, tiltmeters or strainmeters beneath the telescope pier, and nearby borehole or groundwater monitoring can all provide independent constraints on local vertical motion, hydrological loading, and small-scale mass redistribution at the site. In the observation model these data inform priors on the station coordinates and local gravity field and help disentangle truly global signatures in $R_{2\mathrm{w}}(t)$ and $v_r(t)$ from slowly varying, site-specific effects. A superconducting gravimeter in particular can track local gravity changes at the $10^{-11}\,g$ level on timescales from minutes to years, which is directly relevant for separating long-period tidal signals and hydrological loading from the relativistic and interior-physics parameters entering $R_{\rm geom}(t;\boldsymbol{\theta})$ in Eq.~\eqref{eq:Rtot}. While such ancillary sensors are not required to meet the single-window precision targets in Table~\ref{tab:summary-2w}, they would substantially strengthen the robustness of global fits that combine multi-decade AM--CW LLR with other geophysical data sets and tie the station cleanly into the broader International Terrestrial Reference Frame (ITRF) and Earth-rotation frameworks.

Taken together with the thermal, vibrational, atmospheric, and timing controls described above, this geodetic context ensures that observatory-- and reference--frame systematics enter the analysis primarily through well--characterized priors on $R_{\mathrm{geom}}(t;\boldsymbol{\theta})$, rather than as unmodeled contributions to the per--window covariance $C_{ij}(T)$, keeping the station--level error budget consistent with the allocations adopted in Secs.~\ref{sec:errors}--\ref{sec:perf}.

\subsection{Concept of operations}
\label{sec:conops}

The following concept of operations (CONOPS) is envisioned:
\begin{itemize}
\item \emph{Point and acquire:} slew the telescope to the nominal reflector position; acquire at low power; close pointing by maximizing the return amplitude in a short integration (sub--second). 

\item \emph{Internal calibration:} run the internal reference continuously; time--align and subtract its phase per tone from the lunar phases; monitor the residual as a function of time and temperature.  Operationally, the internal reference defines the zero of $R_{\mathrm{inst}}(t)$ and transfers most of the internal optical and RF path fluctuations into a common-mode phase that is removed before forming the estimators in Sec.~\ref{sec:model}. The residual instrumental contribution entering ${\bf C}_{\mathrm{inst}}(T)$ and (\ref{eq:budget}) is therefore dominated by imperfect tracking of this common mode (finite loop bandwidth, thermal gradients, and calibration offsets); these terms are explicitly allocated at the $\sim 40~\mu\mathrm{m}$ level in Sec.~\ref{sec:instrum-er}.

\item \emph{Science windows:} transmit the full multi--tone AM pattern; unwrap $\phi(t)$ for each tone. For windows $T = 10, 30, 100$~s, compute the mean phase on the precision tone $\rightarrow$ two--way range and the linear phase slope $\rightarrow$ one--way range--rate; integrate across multiple windows to form normal points suitable for LLR analysis. In parallel, record $f_m(t)$ with sufficient resolution to reconstruct the deterministic component of the modulation-frequency history across the $\sim 2.56$~s round trip, so that the range--rate bias associated with frequency slew [Eq.~(\ref{eq:slew})] can be removed in post-processing.

\item \emph{Differential mode:} interleave CCRs A/B on a cadence short compared with slow differential-delay drift and internal-reference drift, while treating millisecond-scale optical turbulence statistically through \(\sigma_{\Delta R,\rm atm}(T,\theta)\). In a strictly monostatic single-field implementation, the transmit and receive schedule must account for the \(\tau_{2\mathrm{w}}\simeq2.56~\mathrm{s}\) lunar round trip: photons transmitted toward target A return after the telescope may already have been commanded toward target B. An effective 1--2~s A/B cadence is therefore a hardware and timing requirement, not an assumption. It requires either a receive field and steering model that preserves the previous target's return, a near-monostatic split transmit/receive geometry, or a multiplexed pointing/timing sequence that assigns each return to the correct transmitted tone and target. The pipeline forms \(\Delta R\) and \(\Delta \dot R\) on the same analysis grid and carries the photon lower bound of Eq.~\eqref{eq:diff-photon-floor} explicitly in the covariance.

\item \emph{Quality gates:} enforce elevation $e \ge 30^\circ$; require minimum tone SNR on the precision tone; constrain $|\dot f_m|$ via direct measurement; monitor internal--reference residuals; reject windows exhibiting obvious cycle slips, dropouts, or strong background excursions. These gates enforce the per-window bounds on SNR, oscillator drift, internal-reference stability, and elevation that underlie the scalar error budget in Eq.~(\ref{eq:budget}) and are made explicit as implementation targets in Sec.~\ref{sec:gates} and Table~\ref{tab:gates}.

\end{itemize}

Each observing block thus produces a well--defined set of data products:
(i) raw phase time series $\phi^{(j)}(t)$ per tone and CCR;
(ii) internal reference phase $\phi_{\rm inst}(t)$;
(iii) the recorded modulation--frequency history $f_m(t)$;
(iv) environmental telemetry (temperature, pressure, $r_0$, $\tau_0$, $\theta_0$, wind); and
(v) quality flags identifying windows with low SNR, suspected cycle slips, or abnormal instrument behavior.
From these, the reduction pipeline constructs calibrated observables $\widehat{R}_{2{\rm w}}(T)$, $\widehat{v}_r(T)$, $\Delta\widehat{R}_{2{\rm w}}(T)$, and $\Delta\widehat{v}_r(T)$ with associated covariance estimates derived from the equations in Secs.~\ref{sec:model} and~\ref{sec:errors}.
The implementation gates in Sec.~\ref{sec:gates} are chosen such that each hardware or environmental parameter
operates with at least a factor of two margin relative to these allocations. Taken together, these operational steps enforce the variance allocations for $\Cshot(T)$, $\Catm(T)$, $\Cinst(T)$, $\Cosc(T)$, and $\Cnl(T)$ of Sec.~\ref{sec:model} on each integration window $T$.

\section{Conclusions} 
\label{sec:concl}

In this work we developed a complete amplitude-modulated continuous-wave (AM--CW) metrology framework for high--power lunar laser ranging (LLR) built around RF phase measurements on a bright optical carrier  \cite{Turyshev:CW-LLR:2025}. Starting from an explicit model for the modulated transmitted and received signals, we defined two--way range $R_{2\mathrm{w}}(T)$ and one--way line--of--sight range--rate $v_r(T)$ as joint estimators on the RF envelope phase and its slope over an integration window $T$. The observables and their differential counterparts are represented by a covariance matrix ${\bf C}(T)$ that is decomposed into photon, atmospheric, instrumental, oscillator, and nonlinearity contributions. This structure connects the AM--CW measurement model directly to global parameter estimation for the Earth--Moon system.

Using the photon-return regimes inherited from the high-power CW link analysis, we find that the photon-statistical floor at \(f_m=1~\mathrm{GHz}\) is case-dependent. On \(T\simeq100~\mathrm{s}\) windows it is
\(81\)--\(135~\mu\mathrm{m}\) for the generic high-power case,
\(30\)--\(55~\mu\mathrm{m}\) for the dedicated AM--CW case, and \(22\)--\(30~\mu\mathrm{m}\) for photon-rich operation. In the dedicated design case, a representative allocation
\[
\sigma_{R,\mathrm{shot}} \simeq 30~\mu\mathrm{m},\quad
\sigma_{R,\mathrm{atm}} \simeq 60~\mu\mathrm{m},\quad
\sigma_{R,\mathrm{inst}} \simeq 40~\mu\mathrm{m}
\]
implies a total absolute range precision \(\sigma_R\simeq0.08~\mathrm{mm}\)
in root-sum-square. For range-rate, the corrected slope estimator gives sub-\(\mu\mathrm{m\,s^{-1}}\) sensitivity on several-hundred-second Case~B windows, or on shorter windows only in photon-rich operation.

To connect these formal error budgets to realistic stations, we grouped the link, turbulence, and metrology assumptions into three representative operating regimes. Case~A reproduces a generic 1~kW CW implementation on a conventional 1~m observatory platform, with detected photon rates $\dot N_\gamma \simeq (5$--$7)\times 10^3~\mathrm{s}^{-1}$ and total two-way precision in the $0.32$--$0.55$\,mm band, consistent with earlier high-power CW LLR error budgets. Case~B represents the dedicated AM--CW facility developed in this work, with $\dot N_\gamma \sim (3$--$5)\times 10^4~\mathrm{s}^{-1}$ and an absolute two-way precision $\sigma_R \simeq 80~\mu\mathrm{m}$ when atmospheric and instrumental residuals are held near $60~\mu\mathrm{m}$ and $40~\mu\mathrm{m}$, respectively. Case~C is a photon-rich, excellent-seeing regime with $\dot N_\gamma \sim 10^5~\mathrm{s}^{-1}$: once these atmospheric and bench allocations are achieved, increasing the flux mainly provides margin, driving the photon term below $\sim 30~\mu\mathrm{m}$ while leaving the absolute error budget dominated by the remaining $\sim 30$--$80~\mu\mathrm{m}$ atmospheric and $\sim 40~\mu\mathrm{m}$ instrumental contributions. In all three regimes the residual neutral atmosphere is modeled with Kolmogorov statistics and modern mapping functions, yielding $\sim 50$--$150~\mu\mathrm{m}$ absolute and $10$--$50~\mu\mathrm{m}$ differential path fluctuations on $T\sim 10$--$100$~s windows that are explicitly carried in $C_{ij}(T)$ rather than treated as ad hoc margins.

We also quantified the benefits and limits of differential AM--CW LLR between nearby CCRs. Differential operation suppresses common-mode station and atmospheric terms, but it cannot suppress the independent photon noise from the two reflector returns. For equal links, \(\sigma_{\Delta R,\rm shot}=\sqrt{2}\,\sigma_{R,\rm shot}\); consequently, nominal Case~B photon rates imply a \(43\)--\(78~\mu\mathrm{m}\) differential
photon floor at \(T=100~\mathrm{s}\) before atmosphere and instrument terms are added. Including representative differential atmosphere and instrument allocations gives a robust Case~B differential range band of \(\sim45\)--\(90~\mu\mathrm{m}\), while photon-rich excellent-seeing operation gives \(\sim35\)--\(60~\mu\mathrm{m}\). The \(20~\mu\mathrm{m}\) level is a
stretch target requiring higher detected flux, longer integration, or both.

The covariance decomposition \(C_{ij}(T)\) is translated into concrete implementation gates on station hardware and operations. A hydrogen maser or equivalent ultra-stable reference keeps the oscillator contribution at the
few-micrometer level in range and well below \(1~\mu\mathrm{m\,s^{-1}}\) in range-rate. Internal referencing through a short metrology path removes most optical and RF bench drift, but the residual must be demonstrated through the
instrument PSD and Allan-deviation requirements of
Eqs.~\eqref{eq:inst-psd-range}--\eqref{eq:instrbudget}. Site selection, elevation cuts, and turbulence monitoring bound the atmospheric term, while multi-tone synthetic wavelengths and strict AM-to-PM constraints limit coherent
nonlinear biases.

In summary, high-power AM--CW LLR with kW-class transmitters, GHz-class modulation, and continuous internal metrology provides a technically credible path to \(\sim0.1~\mathrm{mm}\) absolute ranging under favorable site and instrument conditions, and to sub-\(\mu\mathrm{m\,s^{-1}}\) range-rate sensitivity on several-hundred-second windows. The paper is best understood as a metrology and covariance extension of the high-power CW link analysis of \cite{Turyshev:CW-LLR:2025}. Differential AM--CW LLR offers strong common-mode rejection, but its performance must be reported with the \(\sqrt{2}\) photon lower bound and the actual A/B scheduling constraints included in the covariance. The framework presented here provides observation-level covariances suitable for global parameter estimation; the
final science impact on relativistic-gravity, lunar-interior, and
low-frequency gravitational-wave parameters will require a full global covariance analysis including parameter correlations, atmospheric temporal correlations, station systematics, CCR coordinates, and ephemeris uncertainties.

\section*{Acknowledgments} 
The work described here was carried out at the Jet Propulsion Laboratory, California Institute of Technology, Pasadena, California, under a contract with the National Aeronautics and Space Administration.
  
\appendix
\section{Atmospheric turbulence and time averaging}
\label{app:atm_turbulence}

We model residual atmospheric path fluctuations as a zero-mean random process $\delta R_{\rm atm}(t)$ with a Kolmogorov optical-path structure function
\begin{equation}
    D_{R}(\tau)
    \equiv \big\langle \big[ \delta R_{\rm atm}(t+\tau)
    - \delta R_{\rm atm}(t) \big]^2 \big\rangle
    \simeq D_{0}\,\Big(\frac{\tau}{\tau_{0}}\Big)^{5/3},
    \label{eq:DR_def}
\end{equation}
where $\tau_{0}$ is a characteristic coherence time and $D_{0}$ sets the short-time amplitude. For an estimator that averages over an interval $T \gg \tau_{0}$, the relevant quantity for the range covariance ${\bf C}_{\rm atm}(T)$ is the variance of the time-averaged path
\begin{equation}
    \bar{R}_{\rm atm}(T)
    = \frac{1}{T}\int_{0}^{T}\delta R_{\rm atm}(t)\,dt.
\end{equation}
Expressing $\mathrm{Var}[\bar{R}_{\rm atm}(T)]$ in terms of the covariance
$B_{R}(\tau)$ or $D_{R}(\tau)$ and using the power-law
behaviour~\eqref{eq:DR_def} yields the familiar scaling
\begin{equation}
    \sigma_{R,{\rm atm}}^{2}(T)
    \equiv \mathrm{Var}[\bar{R}_{\rm atm}(T)]
    \simeq K_{R}^{2}\,\frac{\tau_{0}}{T},
    \qquad T \gg \tau_{0},
    \label{eq:sigmaR_atm_T}
\end{equation}
where $K_{R}$ is an effective amplitude that absorbs the details of $D_{0}$, elevation, and site-dependent turbulence parameters. In our covariance description we represent this contribution as
\begin{equation}
    {\bf C}_{\rm atm}(T)
    =
    \begin{pmatrix}
        \sigma_{R,{\rm atm}}^{2}(T) & 0 \\
        0 & \sigma_{v_r,{\rm atm}}^{2}(T)
    \end{pmatrix},
    \qquad
    \sigma_{v_r,{\rm atm}}(T)
    \simeq \frac{\sigma_{R,{\rm atm}}(T)}{T},
    \label{eq:C_atm_single}
\end{equation}
where the range-rate variance scaling follows from interpreting $v_r$ as a slope estimated over the same interval $T$.

For differential LLR we require the statistics of the path difference between two closely separated lines of sight, with zenith-angle separation $\theta \ll 1\,\mathrm{rad}$. Let $\delta R_{\rm atm}^{(1)}$ and $\delta R_{\rm atm}^{(2)}$ denote the two paths, and define the differential quantity $\Delta R_{\rm atm} = \delta R_{\rm atm}^{(1)}-\delta R_{\rm atm}^{(2)}$. 

For small separations the spatial correlation is parametrized by an effective differential-delay angular scale \(\theta_0\), so that the differential variance of the time-averaged path obeys
\begin{equation}
    \sigma_{\Delta R,{\rm atm}}^{2}(T,\theta)
    \equiv \mathrm{Var}\big[\overline{\Delta R}_{\rm atm}(T)\big]
    \simeq K_{\Delta R}^{2}
    \left(\frac{\theta}{\theta_{0}}\right)^{5/3}\,
    \frac{\tau_{0}}{T},
    \qquad
    \theta \ll \theta_{0},\; T \gg \tau_{0},
    \label{eq:sigmaDeltaR_atm_T}
\end{equation}
with $K_{\Delta R}$ an effective amplitude that may differ slightly from $K_{R}$. This scaling captures the key behavior: differential atmospheric noise is suppressed by time averaging, through the factor \(\tau_0/T\), and by keeping the reflector separation below the empirical differential-delay angular scale, through the factor \((\theta/\theta_0)^{5/3}\). The parameter \(\theta_0\) is therefore a calibrated path-delay decorrelation scale, not the adaptive-optics wavefront isoplanatic angle. In the differential covariance matrix for the 4-component estimator
\[
  \mathbf{y}_\Delta =
  (\Delta R_{2w},\Delta v_r,R^{\rm ref}_{2w},v^{\rm ref}_r)^T ,
\]
Eq.~\eqref{eq:sigmaDeltaR_atm_T}  determines the \(\Delta R_{2w}\) entry and, through an analogous \(\sigma_{\Delta v_r,\rm atm}(T)\simeq
\sigma_{\Delta R,\rm atm}(T,\theta)/T\), the corresponding differential range-rate term.

For illusration, at a site with $r_0 \simeq 10~\mathrm{cm}$ at 500~nm ($r_0\simeq 25~\mathrm{cm}$ at 1064~nm), $\tau_0 \simeq 5~\mathrm{ms}$, and effective differential-delay angular scale \(\theta_0\simeq1^\circ\), Eq.~\eqref{eq:sigmaR_atm_T} with $K_R$ chosen to match existing mm--class LLR experience yields $\sigma_{R,{\rm atm}}(T)\sim (5$--$10)\times 10^{-5}~\mathrm{m}$ for $T=30$--$100~\mathrm{s}$. For \(\theta\simeq0.05^\circ\), \(T=30\)--\(100~\mathrm{s}\), and an A/B sequence whose timing model correctly assigns the delayed returns, Eq.~\eqref{eq:sigmaDeltaR_atm_T} gives \(\sigma_{\Delta R,\rm atm}(T,\theta)\sim10\)--\(30~\mu\mathrm{m}\) under favorable conditions, consistent with the atmospheric and differential allocations in Table~\ref{tab:summary-2w}. The 1--2 s interleaving cadence controls slow differential drift and target assignment; the millisecond-scale turbulence is already represented statistically by \(\tau_0/T\).

As a cross-check, we have verified the consistency of the analytic scalings with a simple Monte Carlo experiment in which synthetic time series of path fluctuations are generated with a Kolmogorov power spectrum, sampled on representative integration windows, and processed through the same averaging and slope estimators used in the main text. The recovered distributions of range and range--rate uncertainties reproduce the predicted \(\sigma_{R,\mathrm{atm}}(T) \propto T^{-1/2}\) and \(\sigma_{v_r,\mathrm{atm}}(T) \propto T^{-3/2}\) behaviour and fall within the target bands for the adopted site parameters, confirming that the atmospheric contribution to the covariance matrix can be modelled reliably at the design level.

%

\end{document}